\documentclass[prb,superscriptaddress,twocolumn,showpacs]{revtex4}
\usepackage{epsfig}

\begin{document}
\title{Exciton and negative trion dissociation by an external \\ electric field
in vertically coupled quantum dots}
\author{B. Szafran}
\affiliation{Departement Fysica, Universiteit Antwerpen (Campus
Drie Eiken), Universiteitsplein 1, B-2610 Antwerpen, Belgium}
\affiliation{Faculty of Physics and Applied Computer Science, AGH
University of Science and Technology, al. Mickiewicza 30, 30-059
Krak\'ow, Poland}
\author{T. Chwiej}
\affiliation{Departement Fysica, Universiteit Antwerpen (Campus
Drie Eiken), Universiteitsplein 1, B-2610 Antwerpen, Belgium}
\affiliation{Faculty of Physics and Applied Computer Science, AGH
University of Science and Technology, al. Mickiewicza 30, 30-059
Krak\'ow, Poland}
\author{F.M. Peeters}
\affiliation{Departement Fysica, Universiteit Antwerpen (Campus
Drie Eiken), Universiteitsplein 1, B-2610 Antwerpen, Belgium}
\author{S. Bednarek}
\affiliation{Faculty of Physics and Applied Computer Science, AGH
University of Science and Technology, al. Mickiewicza 30, 30-059
Krak\'ow, Poland}
\author{J. Adamowski}
\affiliation{Faculty of Physics and Applied Computer Science, AGH
University of Science and Technology, al. Mickiewicza 30, 30-059
Krak\'ow, Poland}
\author{B. Partoens}
\affiliation{Departement Fysica, Universiteit Antwerpen (Campus
Drie Eiken), Universiteitsplein 1, B-2610 Antwerpen, Belgium}

\begin{abstract}
We study the Stark effect for an exciton confined in a pair of
vertically coupled quantum dots.  A single-band approximation for
the hole and a parabolic lateral confinement potential are adopted
which allows for the separation of the lateral center-of-mass
motion and consequently for an exact numerical solution of the
Schr\"odinger equation. We show that for intermediate tunnel
coupling the external electric field leads to the dissociation of
the exciton via an avoided crossing of bright and dark exciton
energy levels which results in an atypical form of the Stark
shift. The electric-field-induced dissociation of the negative
trion is studied using the approximation of frozen lateral degrees
of freedom. It is shown that in a symmetric system of coupled dots
the trion is more stable against dissociation than the exciton.
For an asymmetric system of coupled dots the trion dissociation is
accompanied by a positive curvature of the recombination energy
line as a function of the electric field.
\end{abstract}
\date{\today}
 \maketitle

\section{Introduction}

Strained self-assembled InAs/GaAs quantum dots grown on subsequent
layers stack spontaneously one above the other\cite{Fa1,Fafard}
forming artificial molecules with spatially extended states due to
the tunnel interdot coupling. The photoluminescence (PL) spectrum
of the coupled dots consists of a number of lines which are blue
or red shifted by the coupling \cite{Fafard,Bayer} depending on
the way the single-particle electron and hole wave functions
contribute to the exciton states in question.\cite{Szafran}
Application of an electric field oriented along the growth
direction offers the possibility of external control of the
strength of the tunnel coupling. Recent experimental
results\cite{Ruth} on the Stark effect for vertically coupled
pairs of nonidentical dots showed the effect of tunnel coupling
through the appearance of avoided crossings between states
localized in different dots. Previously, tunnel-coupling related
Stark shift of the electroabsorption spectra has been observed in
vertical stacks of several quantum dots.\cite{korean}

Stark effect on the exciton states in vertically coupled
self-assembled quantum dots has previously been studied in Refs.
[\cite{Leburton}] and [\cite{Karen}]. An anomaly in the
ground-state Stark shift was found\cite{Leburton} by the ${\bf k}
\cdot {\bf p}$ method accounting for the strain effects and
realistic shapes of the dots. This anomaly consists in deviation
of the ground-state energy line from the usual quadratic
dependence\cite{TCK} on the external field
\begin{equation} E(F)=E(F_0)-p(F-F_0)-\beta (F-F_0)^2, \end{equation} where $F_0$ is the
electric field for which the overlap of the electron and the hole
wave functions is the largest and for which the recombination
energy is maximal, $p$ is the dipole moment and $\beta>0$ -- the
polarizability. The shift calculated\cite{Leburton} for coupled
dots can only be approximated with two parabolas: one for $F<F_0$
and the other for $F>F_0$, amounting in a cusp at $F_0$. Although
this deviation was attributed\cite{Leburton} to the strain
distribution it was shown that such a behaviour can also be
obtained in a single band model of coupled quantum disks
neglecting the strain. \cite{Karen} Actually, as we discuss below
analyzing the Stark shift of the first excited state, this
deviation is due to a near degeneracy of the ground-state around
$F_0$ resulting from the weakness of the hole tunnel coupling. In
the present paper we report on another deviation of the Stark
shift from quadratic form related to the exciton dissociation via
a ground-state anticrossing of a bright state with both carriers
in the same dot and a dark state with separated carriers.

Quantitative
modelling\cite{asm1,asm2,asm3,asm4,asm5,asm6,asm7,asm8,asm9,asm10}
of single quantum dots requires taking into account the valence
band mixing, the gradient in the indium distribution, strain
effects, and confinement geometry which are very different for
quantum dots fabricated at various laboratories. In this paper we
present a qualitative study of the effects of the external
electric field on the {\it interdot} tunnel coupling visible in
the Stark shifts of the bright energy levels, which should be
universal for various types of coupled dots. In particular we
focus on the effect of the electron-hole interaction which was
neglected\cite{Leburton} or treated in an approximate
manner\cite{Karen} in previous work. For a single quantum dot the
Coulomb interaction may have a small effect on the Stark shift
since the interaction energy only weakly changes with the small
displacement of the electron and hole wave functions inside a
single dot. On the other hand the role of the interaction for the
Stark effect in coupled dots is essential since the effect of the
external field on the exciton consists in breaking the
electron-hole binding and segregation of carriers into different
dots.

In the present work we use a simple model potential\cite{troiani}
with a square quantum well for the vertical confinement and
parabolic lateral confinement adopting the single band
approximation for the hole. Due to the applied idealizations the
model is exactly solvable. Our results fully account for the
interparticle correlations due to the Coulomb interaction and
cover also the excited states.

A recent experiment\cite{Ruth} on the Stark effect in a vertically
coupled system of quantum dots was performed on a charge tunable
structure, similar to the one used in studies of negatively
charged excitons.\cite{warburton} A spectacular change in the
spectrum was observed,\cite{Ruth} when an electron was trapped in
the dot closer to the electron reservoir. Namely, a sudden drop of
the recombination energy and an unexplained positive curvature of
the recombination line as a function of the electric field was
observed.\cite{Ruth} This observation motivated us to look at the
Stark effect for the negatively charged trion. For the negative
trion we apply the approximation that the lateral degrees of
freedom are frozen. The validity of this approximation is first
verified for the Stark shift of the exciton energy levels. In
nanostructures the trion binding energies with respect to the
dissociation into an exciton and a free electron are considerably
increased.\cite{stebe} However, the trion binding energy is
usually substantially smaller than the exciton binding energy. We
report here that for a symmetric system of vertically coupled
quantum dots the trion is {\it more} stable for dissociation by
the external electric field than the exciton. The study of the
dissociation mechanism shows, that for the pair of identical dots
the trion is dissociated into a pair of electrons confined in one
dot and a hole in the other. Only for the asymmetric system of
coupled dots a dissociation into an exciton and a free electron is
obtained as an intermediate step before the final separation of
the hole from the two electrons. In this case, the trion is more
easily dissociated than the exciton. The positive curvature of the
recombination energy as a function of the electric field is
obtained for the trion ionization process into an exciton and a
free electron.

Previously, trions in vertically coupled dots were studied in the
absence of the external field\cite{ss} and neglecting tunnel
coupling between the dots.\cite{anisimo}

This papers is organized as follows, the next section contains the
description of the theoretical approach, the results are given in
Section III, their discussion is presented in Section IV. Section
V is devoted to the summary and conclusions.

\section{Theory} We assume a parabolic lateral confinement
potential with equal electron and hole confinement energy ($\hbar
\omega$). Vertical confinement for the electron ($V_e(z_e)$) and
the hole ($V_h(z_h)$) is taken as double well potentials of depth
$V_e^0$ for the electron and $V_h^0$ for the hole and of width
$w=6$ nm separated by a barrier of thickness $b$. Isolated quantum
dots may possess a built-in strain-induced electric field pushing
the hole to the top of the dot as found in the photocurrent
measurements of the Stark effect on buried quantum dots.
\cite{Fry} However, in coupled quantum dots the built-in electric
field has the opposite orientation.\cite{Ruth} Therefore, this
intrinsic electric field is neglected in the present calculations
(in fact, such a build in electric field can also be interpreted
as a shift of our applied field). For self-assembled quantum dots
the assumption of harmonic lateral confinement is not valid,
however it should not essentially modify the susceptibility of the
carriers to the electric field oriented vertically.

In the present model the Hamiltonian of the system can be written
as
\begin{eqnarray}
H=&-\frac{\hbar^2}{2 m_e}\nabla^2_e-\frac{\hbar^2}{2
m_h}\nabla^2_h
+\frac{m_e\omega^2}{2}\rho^2_{e}+\frac{m_h\omega^2}{2}\rho^2_{h}
+V_e(z_e)\nonumber \\
&  +V_h(z_h)
-\frac{e^2}{4\pi\epsilon\epsilon_0r_{eh}}+e\Phi(z_e)-e\Phi(z_h) ,
\end{eqnarray}
where $\rho_e^2=x_e^2+y_e^2$, $(x_e,y_e,z_e)$ and $(x_h,y_h,z_h)$
are the position vectors of electron and the hole, respectively.
$r_{eh}$ is the electron hole distance, $m_e$ ($m_h$) is the
electron (hole) effective band mass, $\epsilon$ is the dielectric
constant, and $\Phi(z)$ is the potential of the external electric
field taken as
\begin{equation}
\Phi(z)=\left\{\begin{tabular}{ccc} $F z_{max}$ & for&  $z_{max}\le z$ \\
$F z$ &for& $z_{min} < z < z_{max}$ \\ $F z_{min}$ & for& $z\le
z_{min}$
\end{tabular} \right.,
\end{equation}
where $F$ is the value of the electric field assumed to be uniform
between $z_{min}$ and $z_{max}$ (which can be identified as the
positions of the electrodes). In the calculations we leave a space
of 10 nm between the dots and the points $z_{min}$ and $z_{max}$
beyond which the electric field is assumed to be zero.

The model of the coupled quantum dots used in this paper was
previously applied\cite{troiani} to describe the exciton coupling
between dots in the absence of an external electric field. The
authors\cite{troiani} used the configuration interaction scheme to
account for the lateral correlations between the electron and the
hole. The configuration interaction approach for the electron-hole
systems is computationally much more challenging than for the
electron systems due to its slow convergence.\cite{halonen}
Therefore, in this paper we will make explicit use of the lateral
separability of the center of mass. After introduction of the
lateral relative ${\mbox{\boldmath $\rho$}}_{eh}
=(x_e-x_h,y_e-y_h)$ and lateral center-of-mass ${\mbox{\boldmath
$\rho$}}_{cm}=(m_ex_e+m_hx_h,m_ey_e+m_hy_h)/M$ coordinates, the
Hamiltonian can be expresses as a sum of the lateral
center-of-mass Hamiltonian ($H_{cm}$) and the Hamiltonian for the
relative lateral-- and the single-particle vertical-- motion
($H_{rv}$), which are given by
\begin{equation} H_{cm}=-\frac{\hbar^2}{2M} \nabla^2_{\rho_{cm}}
+\frac{M\omega^2}{2}\rho_{cm}^2
\end{equation}
and
\begin{eqnarray}
H_{rv}  =&-\frac{\hbar^2}{2\mu}\nabla^2_{\rho_{eh}}
-\frac{\hbar^2}{2m_e}\frac{\partial^2}{{\partial}z_e^2}
-\frac{\hbar^2}{2m_h}\frac{\partial^2}{{\partial}z_h^2}
+\frac{\mu\omega^2}{2}\rho^2_{eh} +V_e(z_e)\nonumber \\ &+V_h(z_h)
 -\frac{e^2}{4\pi\epsilon\epsilon_0r_{eh}}+e\Phi(z_e)-e\Phi(z_h) \;,
\label{rel}
\end{eqnarray}
with $M=m_e+m_h$, $\mu=m_e m_h/(m_e+m_h)$,  $\nabla^2_\rho$ stands
for the Laplacian in the $x-y$ plane. The exciton wave function
can be written as
\begin{equation}
\Psi({\bf r_e},{\bf r_h})=\chi(\mbox{\boldmath
$\rho$}_{eh},z_e,z_h)\psi_{cm}(\mbox{\boldmath $\rho$}_{cm}),
\end{equation}
where $\chi$ and $\psi_{cm}$ are the eigenfunctions of the
$H_{rv}$ and the $H_{cm}$ Hamiltonians, respectively. Functions
$\psi_{cm}$ are simply the eigenfunctions of a two-dimensional
harmonic oscillator.

The eigenstates of Hamiltonian (\ref{rel}) have definite $z$
component of total angular momentum and for $F=0$ also have
definite parity with respect to a change of sign of the $z$
coordinates.\cite{Szafran} The absorption/recombination
probability for state $\mu$ is proportional to the integral
\begin{eqnarray} p_\mu&=&\left |\int d^6{\bf r} \Psi_{\mu}({\bf
r_e},{\bf r_h})\delta^3({\bf r_e}-{\bf r_h}) \right |^2 \nonumber
\\&=&\left |\int dx_edy_e \psi_{cm}(x_e,y_e)\int dz_e \chi_{\mu}(0,z_e,z_e)\right|^2.
\label{rp}\end{eqnarray} In the present paper, we consider only
states whose symmetry does not prevent them to be bright, i.e.
states in which {\it both} the relative $\chi$ and the center of
mass $\psi_{cm}$ eigenstates possess zero angular momentum. In the
following we show and discuss only results for states in which the
center of mass is in the ground state. The spectrum with
$s$-symmetry center-of-mass excitations is simply a replica of the
spectrum corresponding to the ground state of the center of mass
shifted by the energy $2\hbar \omega$. The recombination
probabilities for the states corresponding to zero angular
momentum center-of-mass excitations are {\it exactly} equal to the
corresponding states with the ground-state center of mass, since
integrals of all the $s$ type wave functions of a two-dimensional
harmonic oscillator are equal, which is due to a property of
Laguerre polynomials. For potentials, in which the parity is a
good quantum number, i.e. for identical quantum dots without an
external field, we consider only states of even parity, the odd
parity states being dark.

The eigenfunctions $\chi$ of Hamiltonian (\ref{rel}) are
calculated on a three-dimensional finite-difference mesh with the
imaginary time technique.\cite{ktd} We use the material parameters
for an $\mathrm{In_x Ga_{1-x}As}$ quantum dot embedded in a GaAs
matrix with a uniform concentration of indium in the quantum dot
$x=0.66$.\cite{Szafran} We take the following parameters for the
alloyed quantum dot material $\epsilon=12.5$, $m_e=0.037 m_0$,
$m_h=0.45 m_0$, where $m_0$ is the free electron mass,
$V_e^0=-0.508$ eV, $V_h^0=-0.218$ eV, and we take for the lateral
confinement $\hbar \omega=20$ meV. We note, that in the limit of
$\hbar \omega=0$ the present problem reduces to the Stark effect
for an exciton in coupled quantum wells.\cite{Golub}

For a particle of mass $m$ confined in a harmonic oscillator
potential of energy $\hbar \omega_0$ the localization radius
defined as the square root of the expectation value of $x^2+y^2$
is equal to $\sqrt{\hbar/m \omega_0}$. For the assumed
center-of-mass separation the hole is therefore more strongly
localized than the electron by a factor of $\sqrt{m_h/m_e}$. In
InAs/GaAs quantum dots the hole confinement is stronger than the
electron confinement which is due to the finite quantum well
effect\cite{Wojs} and the electron-hole interaction which
localizes the heavy hole much more strongly than the light
electron. In Fig. 7 we show that a change in the strength of the
hole and electron lateral confinement does not influence the
qualitative features of the spectra in an external electric field.
It merely leads to shifts of the energy levels along the energy
axis.

For the negative trion in quantum dots with a rectangular-well
confinement the effect of a stronger hole localization leads to a
larger electron-hole interaction energy than the electron-electron
interaction energy.\cite{jpcm} This produces a red-shift of the
negative trion recombination line which increases with decreasing
size of the dot and consequently leads to a decrease of the
red-shift due to the tunnel effect in coupled quantum
dots.\cite{ss}  In two-dimensional quantum wells the
experimentally observed\cite{ES} positive and trion recombination
energies for zero-magnetic field are nearly equal, although in
strictly\cite{stebe} two-dimensional confinement significantly
lower recombination energy for the positive trion was predicted.
This effect is explained\cite{ES,filinov,XS} by stronger hole
localization. Therefore, the adopted confinement potential takes
into account the electron-hole interaction
enhancement\cite{ss,jpcm,ES,filinov,XS} with respect to the
electron-electron interaction.

\section{Exciton in vertically coupled dots}

\begin{figure*}[t]
\centerline{\hbox{\epsfxsize=42mm\epsfbox[66 282 456
642]{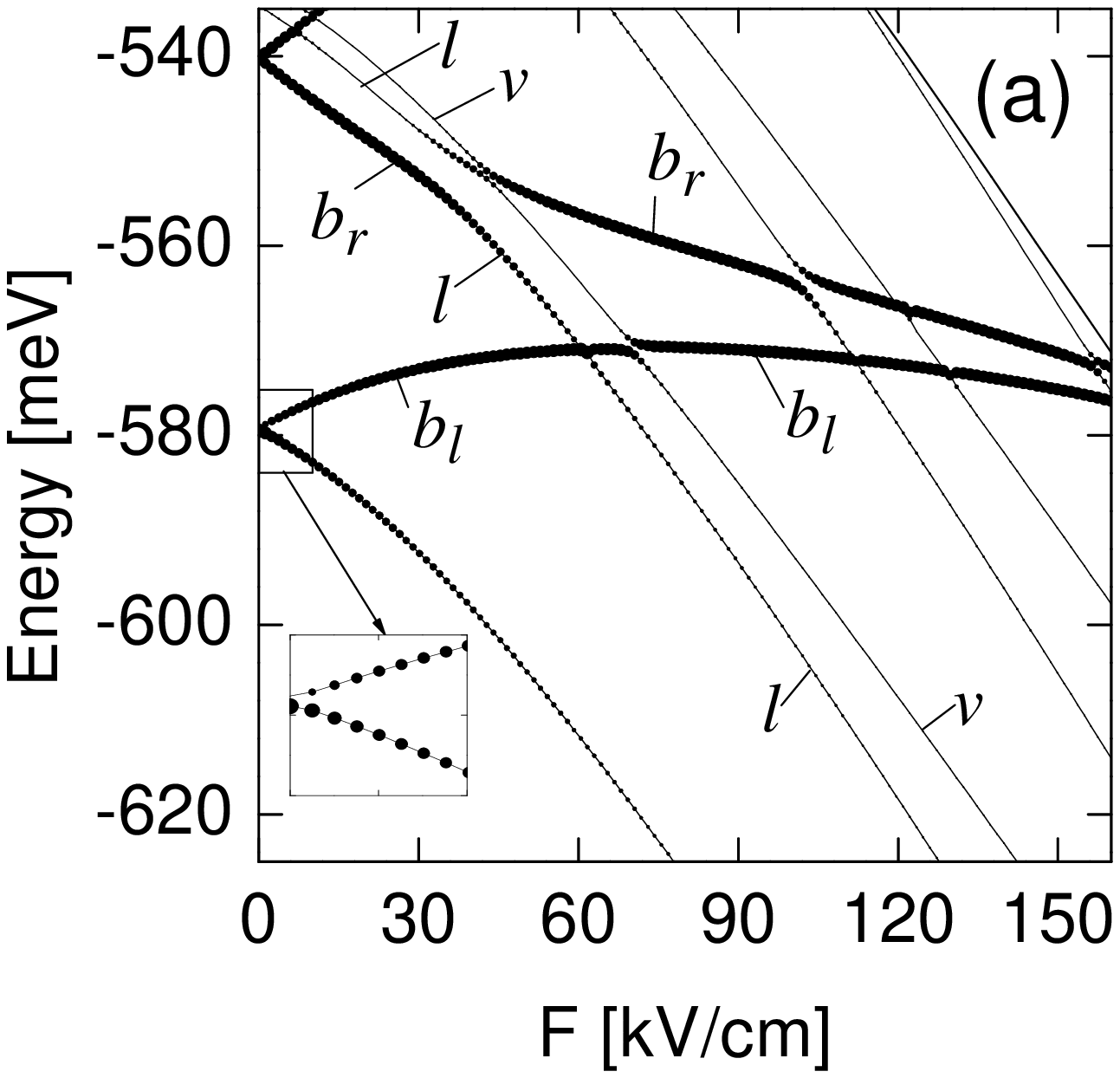}}
    {\epsfxsize=42mm\epsfbox[66 290 443 653]{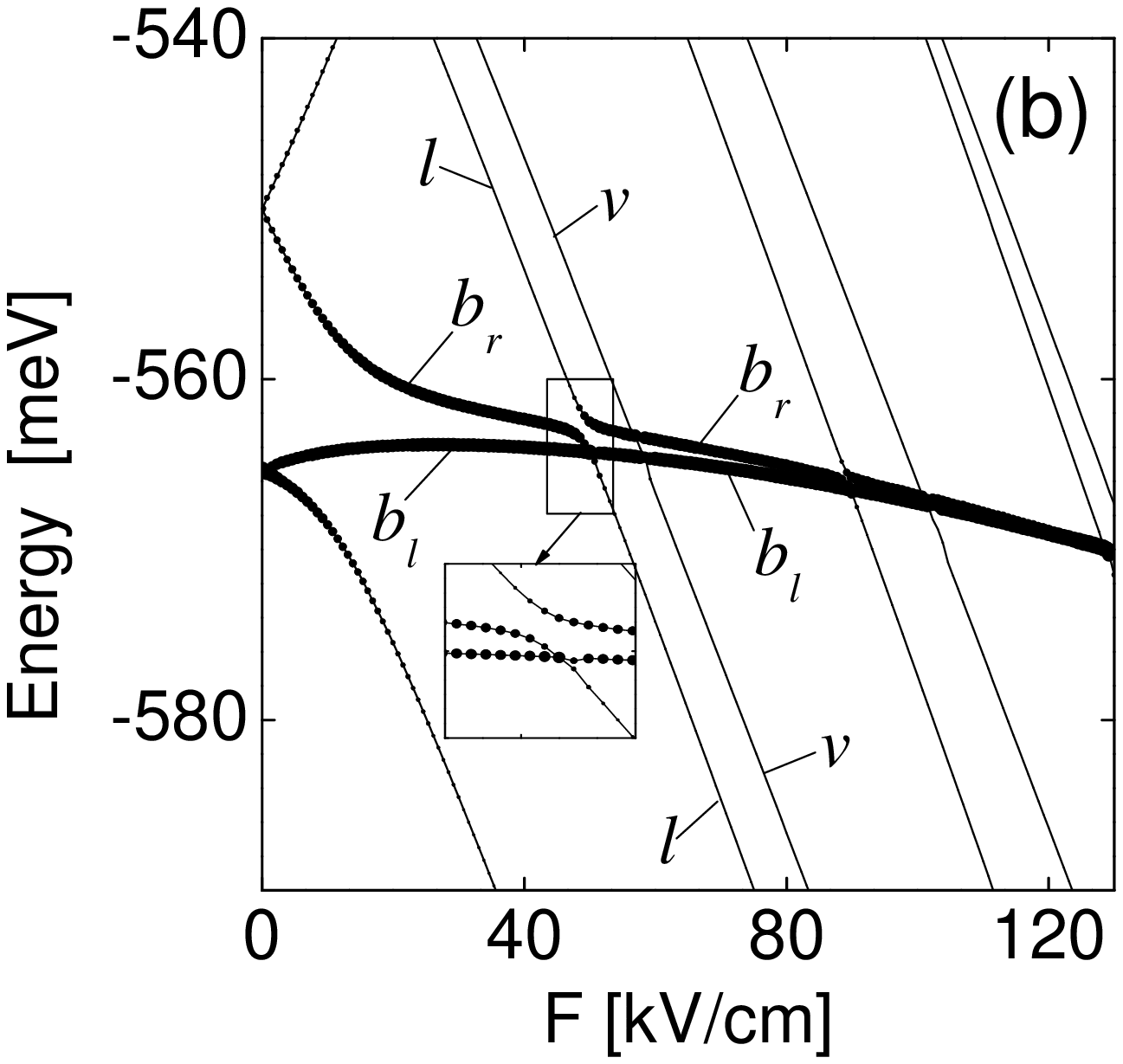}}
{\epsfxsize=42mm\epsfbox[66 290 443 653]{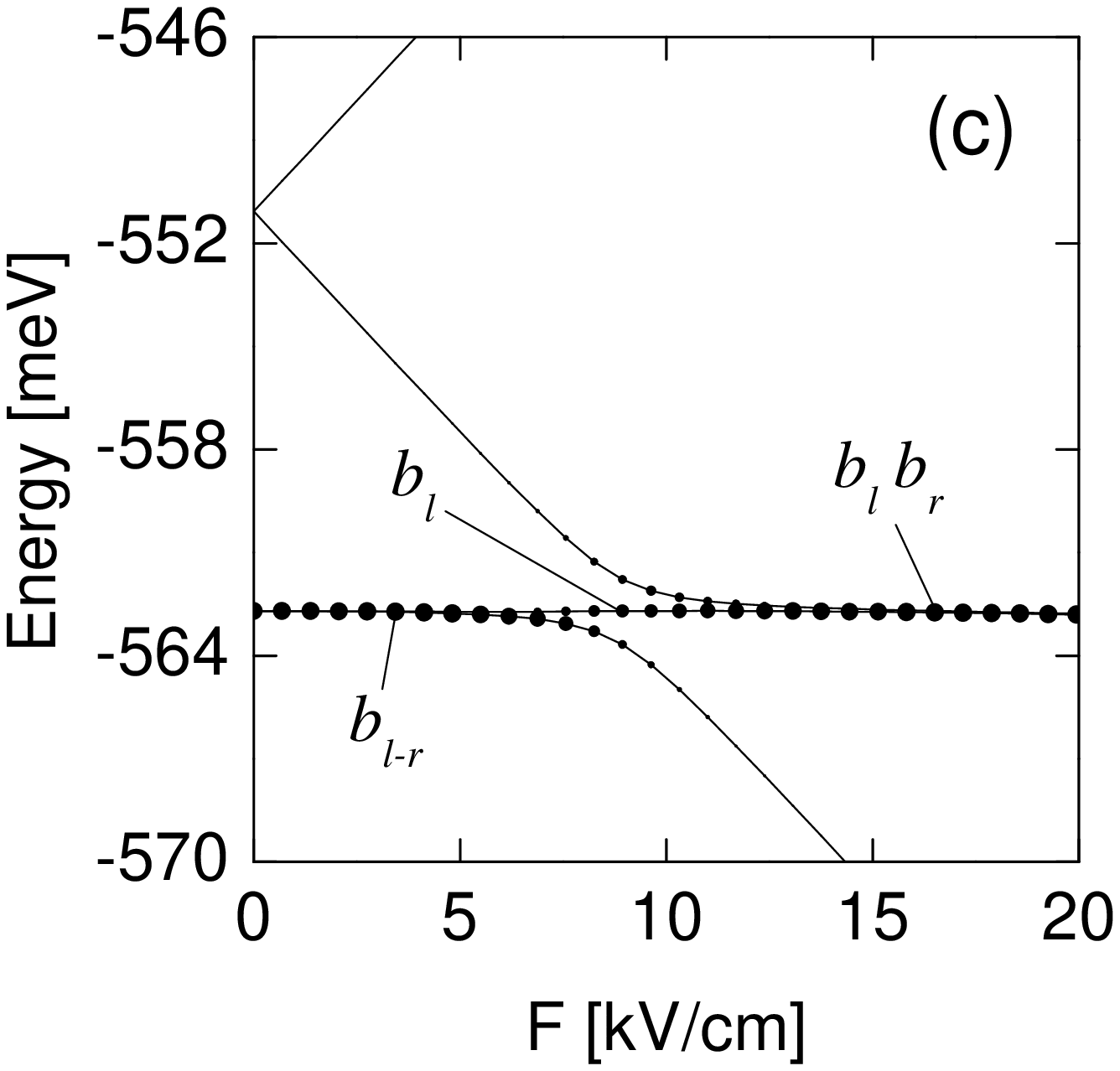}}}

\caption{ Exciton energy spectrum as a function of external
electric field $F$ for barrier thickness $b=2$ nm (a), $b=4$ nm
(b) $b=7$ nm (c). The area of the dots is proportional to the
recombination probability. The insets in (a) and (b) show zooms of
regions marked by rectangles.}
\end{figure*}

\begin{figure*}[t]
\centerline{\hbox{
    \epsfxsize=47.25mm
                \epsfbox[50 35 580 760] {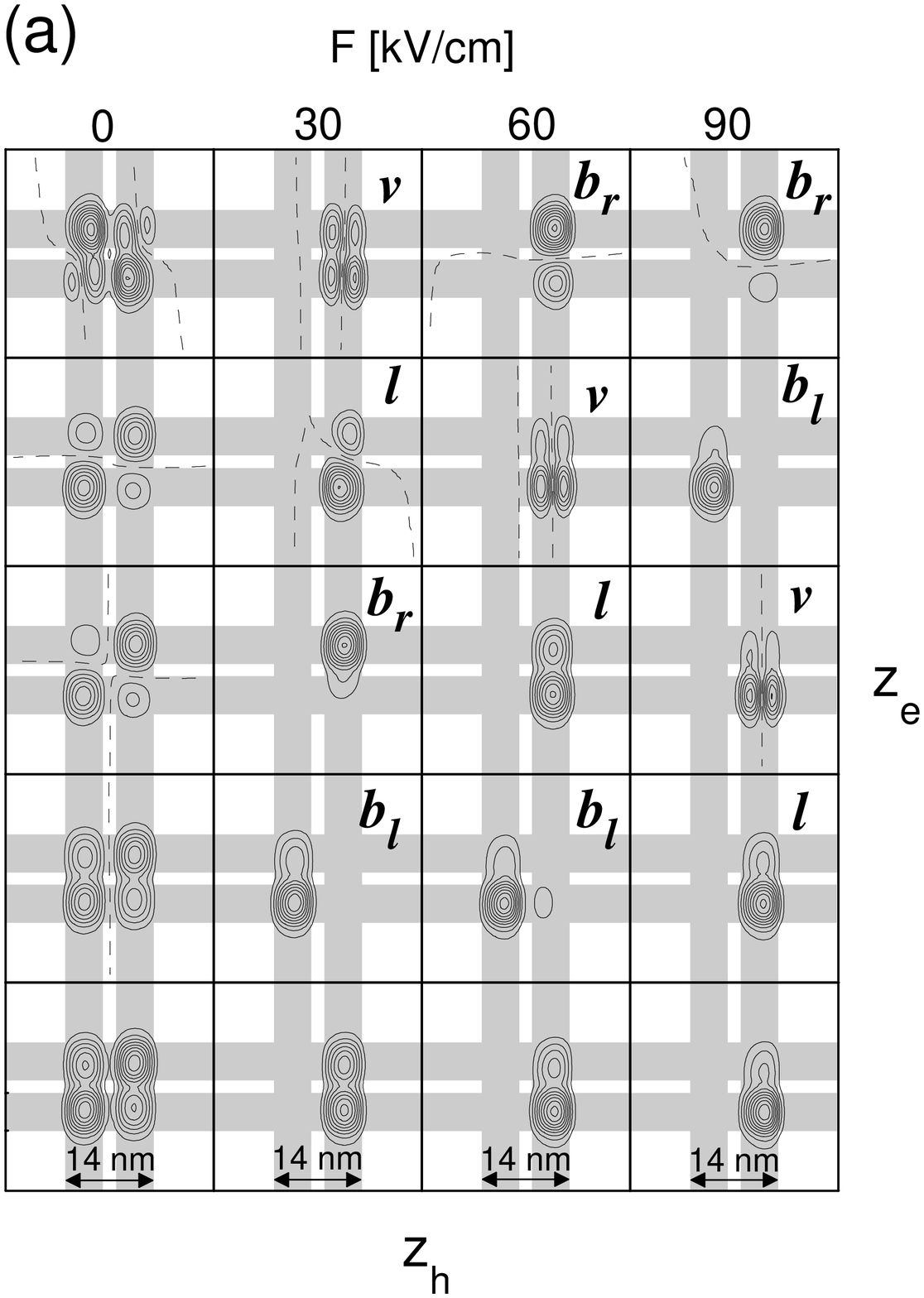}
    \epsfxsize=47.25mm
                \epsfbox[50 35 580 760] {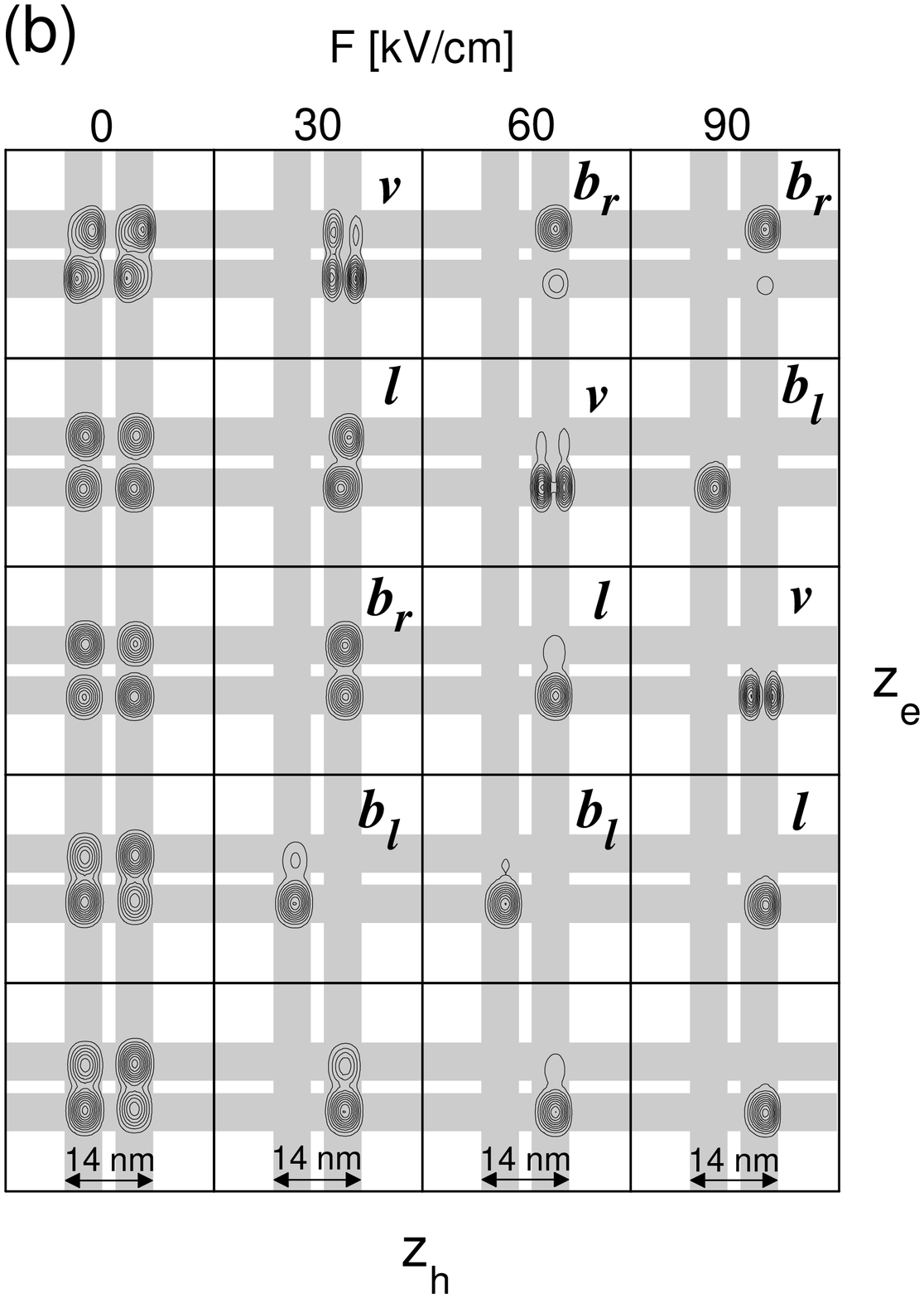}
    \epsfxsize=36.75mm
                \epsfbox[0 35 410 760] {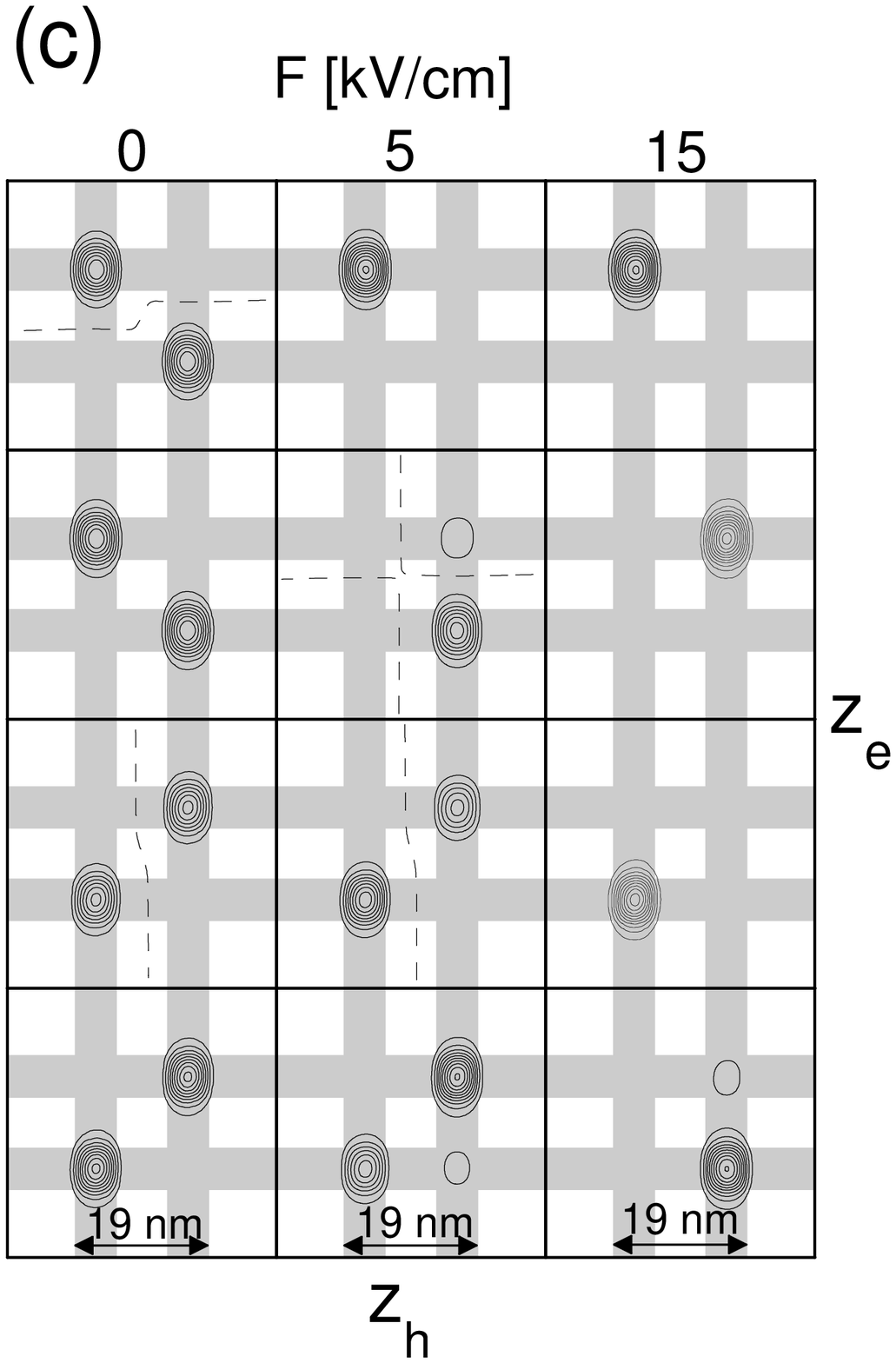}
 }}
\caption{(a,c) contour plots of wave functions at the axis
$\rho=0$ of the system and (b) probability density integrated over
the lateral degrees of freedom $\int_0^\infty d\rho \rho
|\chi(\rho,z_e,z_h)|^2$ for different values of the electric field
for barrier thickness $b=2$ nm (a,b) and $b=7$ nm (c).  Lower
plots correspond to lower energies. Shaded area show the quantum
wells for the electron and for the hole. Dashed line shows the
nodal surface of the wave function.}
\end{figure*}

\subsection{Stark effect}

For $F>0$ the electric field pushes the electron to the left and
the hole to the right dot. The dependence of the energy spectrum
on the external electric field for $b=2$ nm is plotted in Fig.
1(a). At zero electric field the first excited state is of odd
parity and corresponds to the excitation of the hole [see the
inset to Fig. 1(a) -- excitation energy is just 0.25 meV]. The
electric field breaks the parity symmetry of the system and the
excited state becomes optically active [cf. the inset to Fig.
1(a)]. The dependence of the wave functions on the electric field
is displayed in Fig. 2(a). In order to explain the field
dependence of the spectrum we have plotted in Fig. 2(b) the
probability densities integrated over the lateral degrees of
freedom, which gives more accurate information about the
localization of particles than the wave function on the axis
[whose integral over $z_e$ and $z_h$ gives the recombination
probability, cf. Eq. (\ref{rp})]. In the ground state the hole
becomes entirely localized in the right quantum dot for a
relatively weak electric field [see the plots for $F=30$ kV/cm in
Figs. 2(a) and 2(b)]. The ground-state localization of the
electron in the left dot appears at a much higher electric field,
leading eventually to the extinction of the recombination
intensity. In the excited part of the spectrum one observes two
bright energy levels which tend to degeneracy at high electric
field [cf. levels labelled by $b_l$ and $b_r$ in Figs. 1(a), 2(a)
and 2(b)]. In these two energy levels the electron and the hole
occupy the same quantum dot [it is more clearly visible in Fig.
2(b), for the wave function plots presented in Fig. 2(a) this
tendency is apparent only at high electric field, cf. the plots
for the $3^{rd}$ and $4^{th}$ excited states for $F=90$ kV/cm]. In
the bright energy levels marked by $b_l$ the carriers become
localized in the left quantum dot which is favorable for the
electrostatic energy of the electron and unfavorable for the
electrostatic energy of the hole. In the higher bright energy
level marked by $b_r$ the electron and the hole are localized in
the right quantum dot, favorable for the hole and unfavorable for
the electron. The $b_l$ level increases when the electric field is
switched on. On the other hand the $b_r$ energy level decreases
with field. This behavior is due to a reaction of the electron on
the field which is delayed with respect to the reaction of the
hole being more easily localized in one of the dots by the field
[cf. Fig. 2(b)].

Fig. 1(a) shows that the two bright energy levels exhibit avoided
crossings and anticrossings with the dark energy levels for which
the carriers are separated by the electric field in the same way
as in the ground state. The lowest excited dark energy level
[marked by $l$ in Figs. 1(a), 2(a) and 2(b)] corresponds to a
lateral excitation. In the second excited dark energy level
[marked by $v$ in Figs. 1(a), 2(a) and 2(b)] the hole in the right
quantum dot is in a state excited in the vertical direction. For
$b=2$ nm the first anticrossing in the low-energy spectrum appears
between the bright $b_r$ and the dark $l$ energy levels around
$F=40$ kV/cm at about -555 meV. This anticrossing is wide and is
due to the electron tunnel coupling of the left and right dots
(the hole is entirely localized in the right quantum dot in both
states). The dark energy level $l$ goes below the lower bright
energy level $b_l$ via a crossing. A crossing instead of
anticrossing is observed here because in the $b_l$ energy level
the hole is in the other (left) dot. The dark state $v$ with a
hole excitation crosses the $b_r$ level and goes below the $b_l$
level in a very narrow anticrossing.

For weaker tunnel coupling, i.e. for $b=4$ nm [cf. Fig. 1(b)] the
two bright energy levels become degenerate already at about $F=90$
kV/cm. All the avoided crossings become narrower with respect to
the stronger tunnel coupling case of Fig. 1(a). The most
pronounced anticrossing is the one between the $b_r$ and $l$
energy levels, like for $b=2$ nm [cf. Fig. 1(a)]. The curvature of
the degenerate bright energy levels at high electric fields
results from the electric-field-induced deformation of the
electron and hole wave functions within each of the quantum dots.

The most interesting spectrum is obtained for larger barrier
thickness. Fig. 1(c) displays the electric-field dependence of the
exciton energy spectrum for $b=7$ nm. For $F=0$ the twofold
degenerate ground state corresponds to both carriers in the same
quantum dot [cf. Fig. 2(b)], while in the nearly degenerate
excited state the carriers occupy different quantum dots. The
degenerate ground state energy is not affected by the electric
field, since the electrostatic energy gained by the electron is
lost by the hole and {\it vice versa}. The electric-field
dependence of both the split excited energy levels, which
correspond to spatially separated charge carriers, is strictly
linear. This energy level anticrosses the $b_r$ bright energy
level around $F=9$ kV/cm. After the avoided crossing the state
with carriers separated by the external electric field becomes the
ground state. The bright state $b_l$ is not involved in the
anticrossing and its energy is independent of $F$. For larger $b$
the discussed anticrossing becomes narrow and barely visible.

\begin{figure}[t]
{\epsfxsize=50mm
                \epsfbox[142 175 480 512]{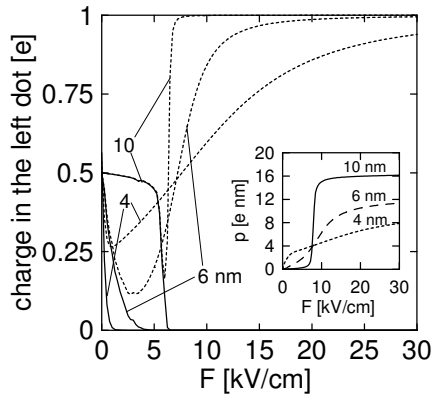}
} \label{fieldx} \caption{Electron (dotted line) and hole (solid
line) charge accumulated in the left quantum dot as function of
the electric field for different barrier thicknesses. Inset shows
the dipole moment as function of the field. } \label{ce}
\end{figure}

Fig. 2(c) for $F=5$ kV/cm shows that in the ground-state the
charge of the hole is considerably shifted to the right dot and
that a part of the electron charge is also transferred to the
right dot. In order to present the movement of the carriers
between the dots in more detail we plotted in Fig. \ref{ce} the
charge accumulated in the left dot as a function of the electric
field for different barrier thicknesses. We see that the
dependence of the hole charge on the external field is monotonous.
On the other hand the electron initially follows the movement of
the hole to the right dot. For $b=10$ nm the electron charge
transferred to the right dot is exactly equal to the hole charge
for $F$ smaller than 6 kV/cm. Up to this field both quantum dots
remain neutral and the dipole moment (see inset to Fig. \ref{ce})
is zero. When both particles become completely localized in
different dots the dipole moment reaches $e(b+w)$.

\subsection{Nonidentical quantum dots}
The confinement potential of vertically stacked dots usually
exhibits asymmetry, which even for identical dots can be induced
by the strain effects.\cite{Leburton} Let us consider the effect
of the asymmetry of the confinement potential on the exciton
spectrum. It was established\cite{Leburton} that for stacked
strained truncated pyramids the ground state of the hole is
completely localized in one of the dots, while the electron
(noninteracting\cite{Leburton} with the hole) still forms bonding
and antibonding states.

\begin{figure}[t]
\centerline{\hbox{
    \epsfxsize=42mm
                \epsfbox[61 277 452 659] {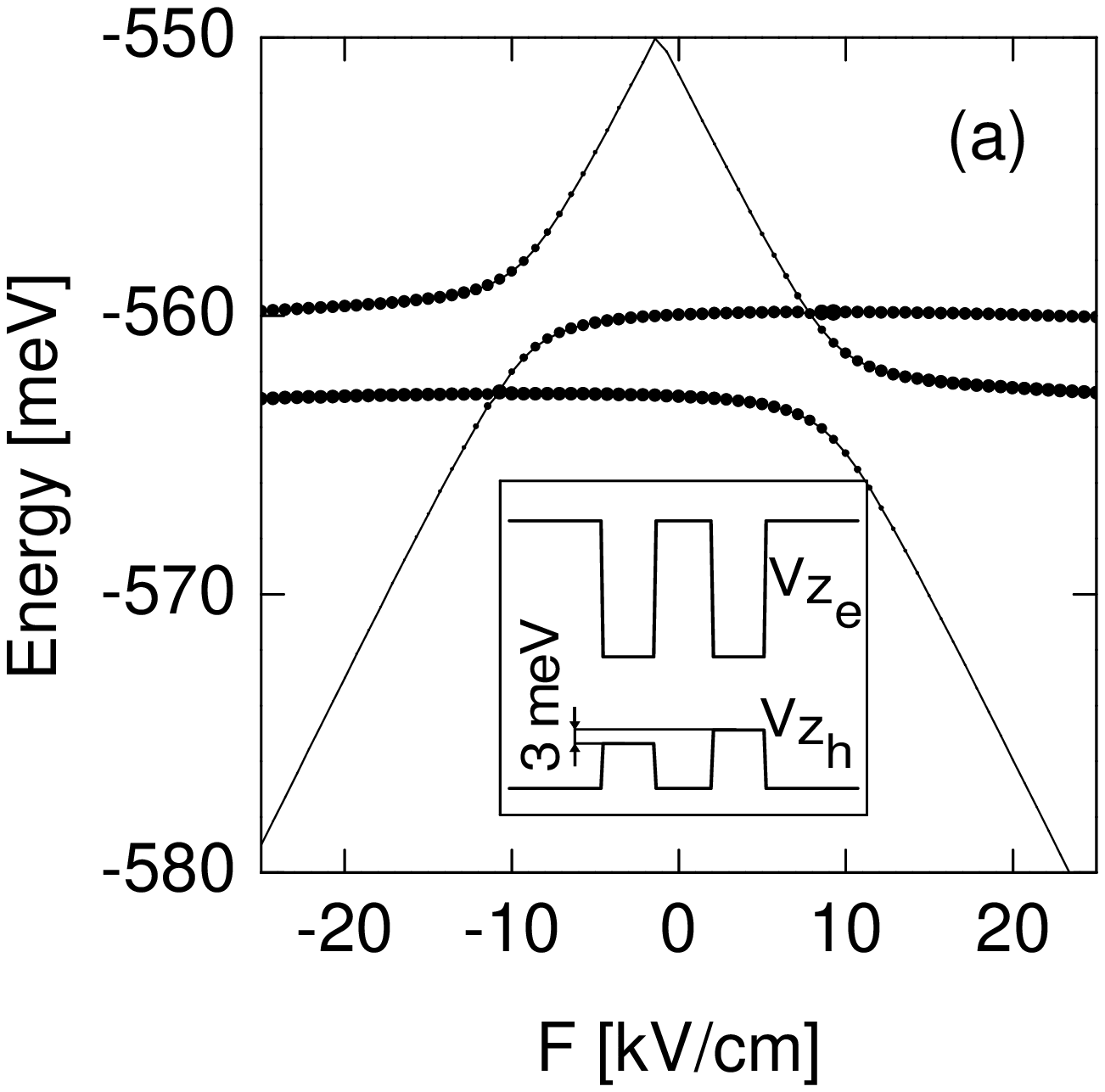}}
    {\epsfxsize=42mm\epsfbox[61 277 452 659]{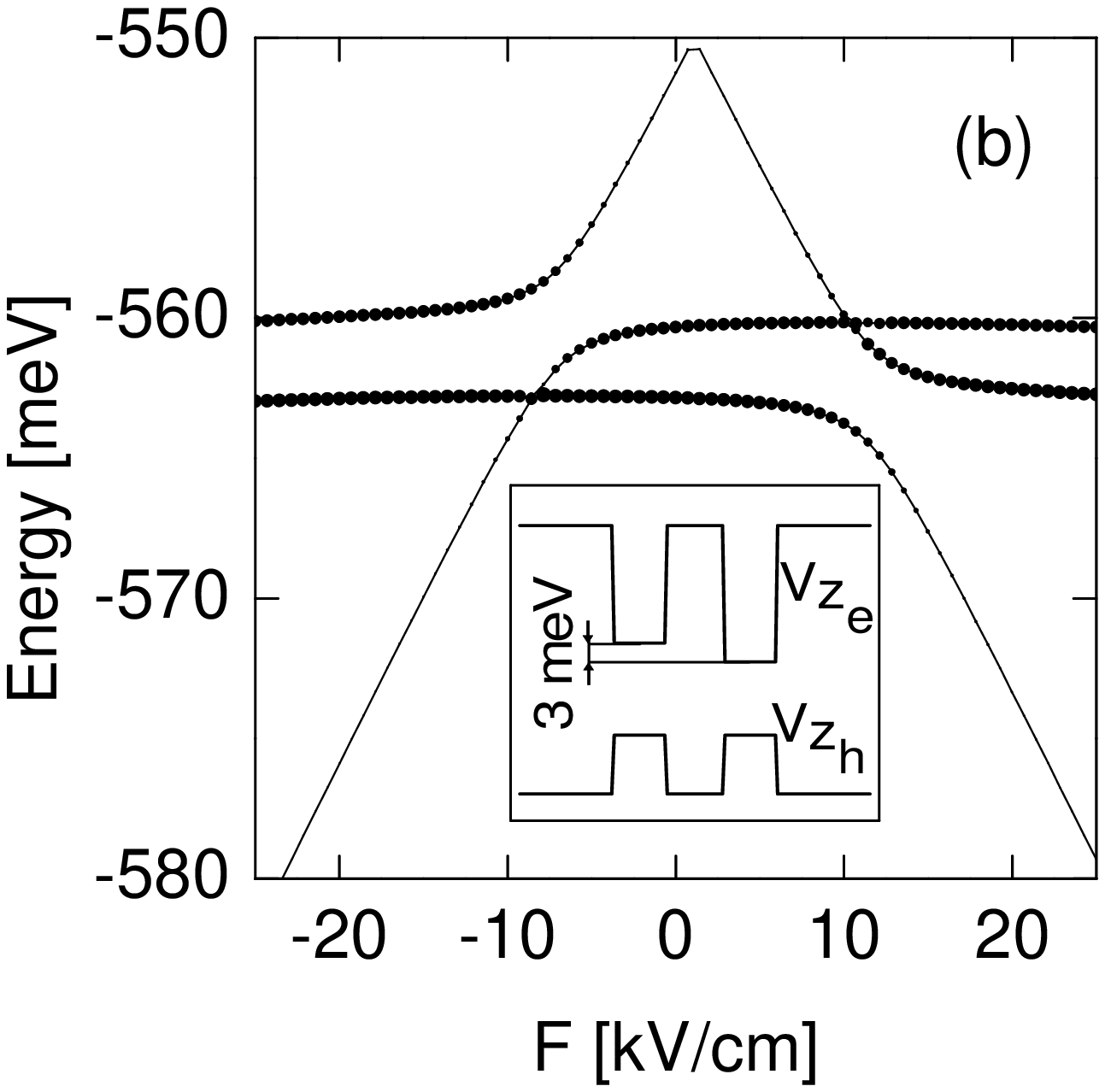}}
    }
\caption{Stark effect for the asymmetric system of quantum dots at
$b=6$ nm. In (a) the electron confinement is symmetric and the
left dot for the hole is shallower by 3 meV. In (b) the hole
confinement is symmetric and the left dot for the electron is
shallower by 3 meV. The insets in (a) and (b) show a schematic
drawing of the vertical confinement for $F=0$. } \label{lhe}
\end{figure}

Here, we simulate this type of localization assuming unequal
depths of the quantum wells for the hole. The effect of the
electric field on the spectrum of asymmetric coupled dots for
$b=6$ nm is presented in Fig. \ref{lhe}(a) for the right dot
deeper by 3 meV for the hole. Two bright energy levels around -563
meV and -560 meV are obtained. In the lower (upper) of energy
levels both the carriers are localized in the deeper (shallower)
of the dots. The lowest dark energy level decreasing linearly in
energy with $F$ has the hole localized in the right dot [cf. Fig.
5(a)] so it crosses the higher bright energy level with both the
carriers in the left dot. The interchange of the energy order of
this dark state with the lower bright energy level appears via an
avoided crossing, since in both these states the hole is localized
in the right (deeper) dot [cf. Fig. 5(a)]. For $F<0$ the hole in
the lowest energy dark state is localized in the shallower of the
dots. For this reason the corresponding energy level anticrosses
the higher bright energy level and crosses the lower one.

Let us now suppose that the left dot is shallower for the electron
(by 3 meV) and that the hole confinement is symmetric. Fig. 4(b)
shows the spectrum for this case. Surprisingly the spectrum for
the electron confinement asymmetry is just shifted by +2.5 kV/cm
with respect to the spectrum for the hole confinement asymmetry
[cf. Fig. 4(a)]. In the lower (upper) of the bright energy levels
the electron stays in the deeper (shallower) of the dots and the
Coulomb interaction binds the hole in the same dot [cf. Fig.
5(b)]. The crossing/anticrossing mechanism is the same as for the
hole confinement asymmetry.

For smaller barrier thickness the anticrossings of the dark and
bright energy levels become wider and as a consequence the region
near $F=0$ in which the two lowest energy levels are nearly
independent of $F$ is narrower. The spectra for the hole asymmetry
for $b=4.5$ nm and $3$ nm are displayed in Fig. 6(a) and (b),
respectively. The two parallel energy levels near $F=0$ observed
for weak tunnel coupling in Fig. 4 are now (see Fig. 6) converted
into a crossing at a small negative $F$. This feature results in
the cusp of the ground-state energy reported
previously\cite{Leburton} for a thin (1.8 nm) interdot barrier.
For the electron asymmetry the spectra are still shifted to higher
values of the field by about 2.5 kV/cm with respect to the hole
asymmetry, like in the weak coupling case of Fig. 4. The crossing
of the bright energy levels still appears at $F<0$. The reason of
this similarity is that in the ground state at $F=0$ the dipole
moment induced by the electron and hole asymmetry is the same in
sign and not very different in size. For small $b$ the electron
charge is smeared over {\it both} dots. If the right dot is deeper
for the electron it binds a {\it larger} part of its charge.
Consequently, the {\it entire} charge of the hole is pulled to the
right dot. On the other hand, for the hole confinement asymmetry
the dot which is deeper for the hole localizes its charge
completely even for small $b$ since the hole tunnel coupling is
negligible. The localization of the hole in the right dot results
in a larger localization of the electron in the right dot. In this
way the asymmetry of the confinement for one particle is
translated into an asymmetry of the potential felt by the other
particle via the Coulomb interaction. Although it is possible
experimentally to determine which of the dots is deeper by looking
at the electric-field dependence of the bright energy levels one
cannot deduce from the $F$-dependence of the exciton energy levels
alone which of the carriers is responsible for the asymmetry.

\begin{figure}[t]
\centerline{\hbox{
    \epsfxsize=30mm
                \epsfbox[175 277 452 820] {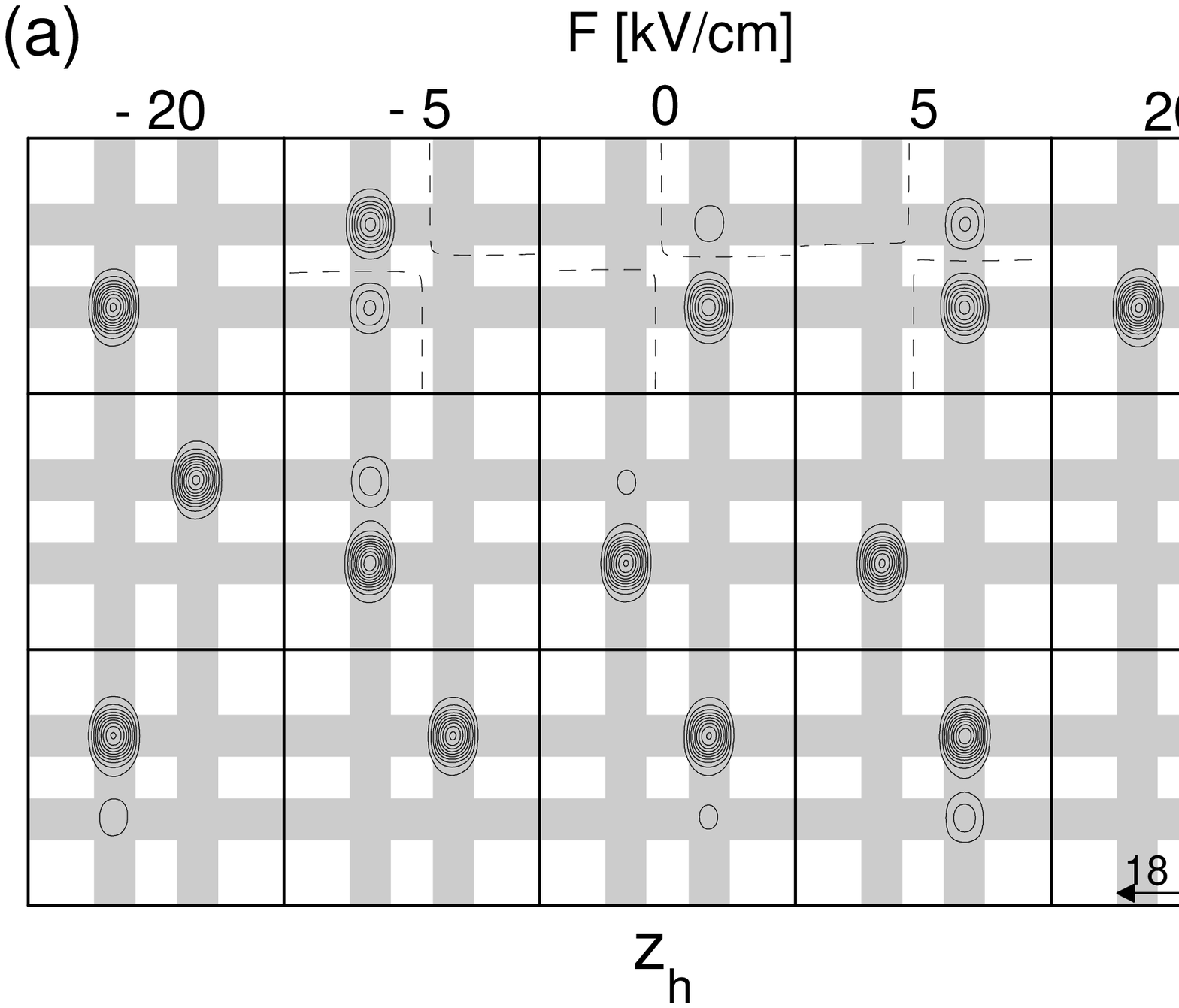}}}
                \centerline{\hbox{
    {\epsfxsize=30mm\epsfbox[190 277 467 820]{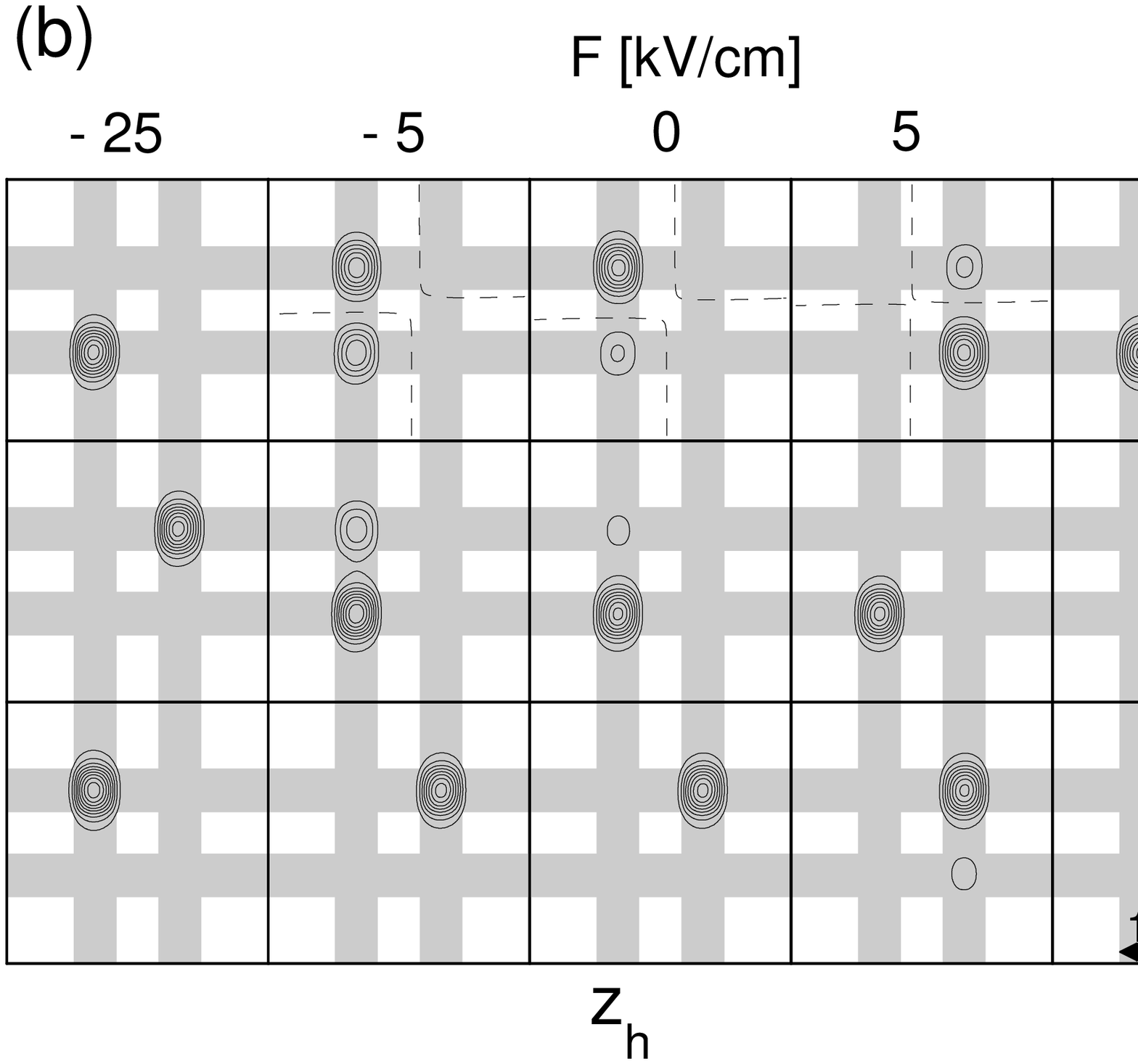}}}
    }
\caption{(a) Contour plots of the wave functions at the axis
$\rho=0$ corresponding to the energy levels shown in Fig. 4(a).
(b) similar as for Fig. 4(b). Higher plots correspond to higher
energies.}
\end{figure}

\begin{figure}[t]
\centerline{\hbox{
    \epsfxsize=42mm
                \epsfbox[61 277 452 659] {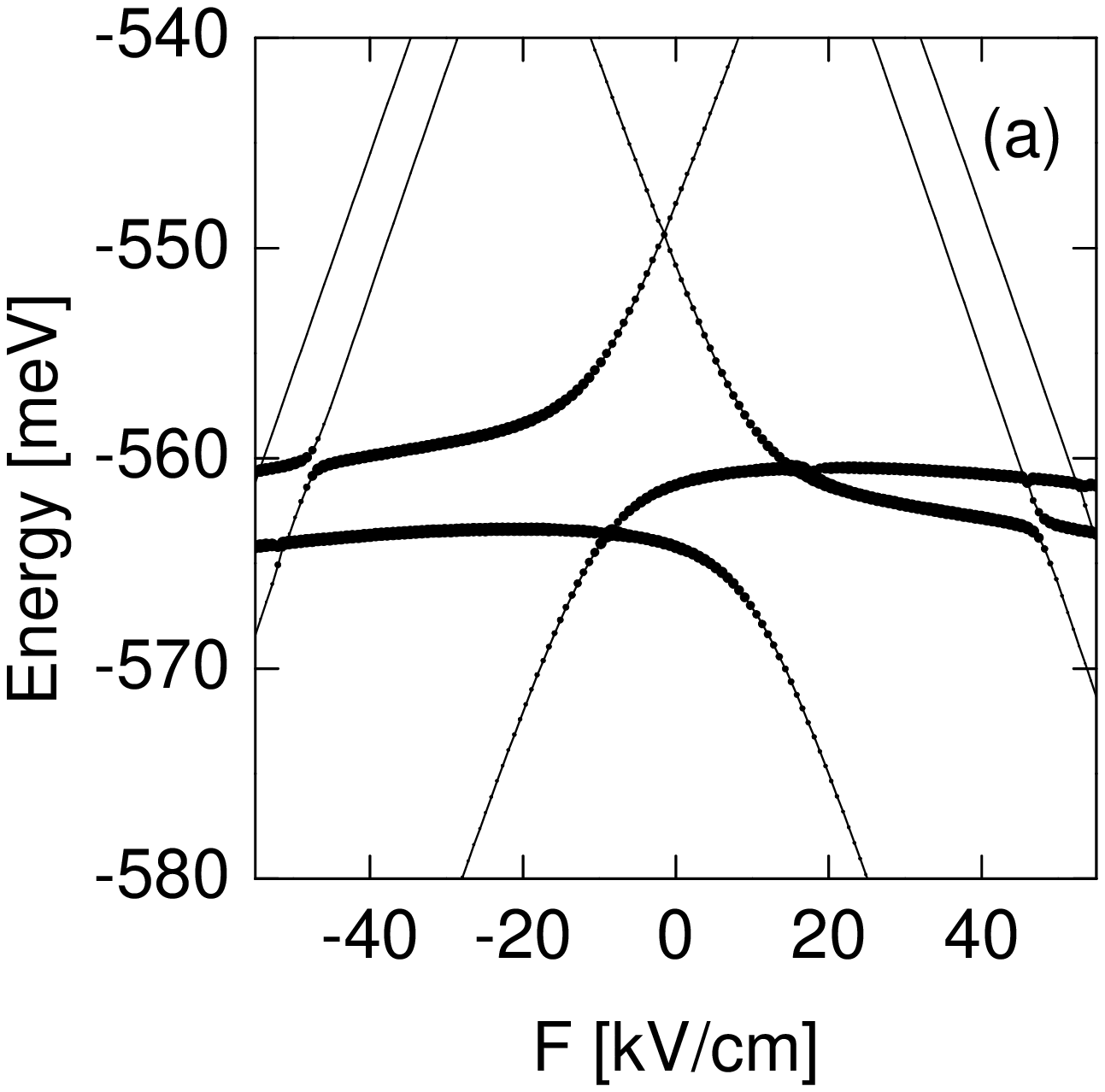}
    {\epsfxsize=42mm\epsfbox[61 277 452 659]{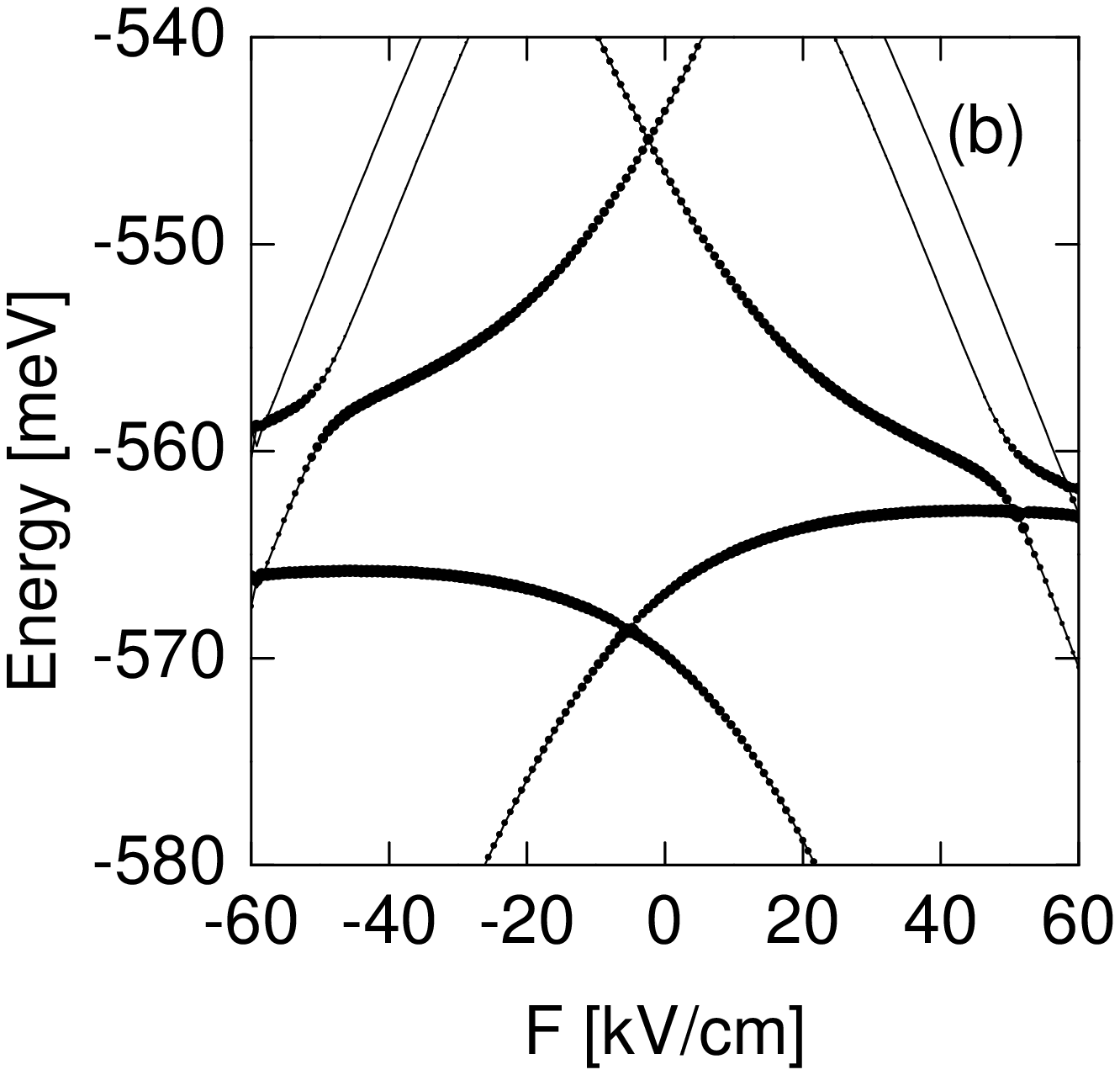}}
    }}
\caption{Stark effect for the asymmetric hole confinement of Fig.
4(a) for $b=4.5$ nm (a) and $b=3$ nm (b). The area of the dots
shows the recombination probability.}
\end{figure}

\subsection{Frozen lateral degrees of freedom}
The exact separability of the center of mass used in the previous
calculations was possible because of the assumption of identical
lateral confinement energies for the electron and the hole. When
the center of mass is not separable\cite{halonen} the exact
calculations become much more complex. However, as long as the
interest of calculations relies in a qualitative description of
the influence of the electric field applied in the growth
direction the actual form of the lateral confinement is not
essential. In this case one may try to integrate out the lateral
degrees of freedom.\cite{1d} Such an adiabatic approximation is
valid for strong lateral confinement, as in the case of
self-assembled quantum dots. Thus we assume that the electron and
hole lateral wave functions can be identified with the
ground-state of the harmonic oscillator. This assumption allows us
to integrate\cite{1d} over the lateral degrees of freedom and
arrive at the effective Hamiltonian for the vertical motion
\begin{eqnarray} H_{\mathrm{eff}}=&
-\frac{\hbar^2}{2m_e}\frac{\partial^2}{\partial z_e^2}
-\frac{\hbar^2}{2m_h}\frac{\partial^2}{\partial
z_h^2}+V_e(z_e)+V_h(z_h)- \nonumber \\ &
V_{\mathrm{eff}}(|z_e-z_h|)+e\Phi(z_e)-e\Phi(z_h)+2\hbar\omega,
\label{hef}
\end{eqnarray}
with $V_{\mathrm{eff}}(z)$ the effective potential\cite{1d} of
one-dimensional interaction given by
\begin{equation} V_{\mathrm{eff}}(z)=\frac{e^2}{4 \pi^{1/2}\epsilon\epsilon_0
l} \mathrm{erfcx}\left(|z|/l \right), \label{pot}
\end{equation} with $l=\sqrt{\frac{\hbar}{\omega}(\frac{1}{m_e}+\frac{1}{m_h})}$.
The solution to the eigenequation of the effective Hamiltonian
(\ref{hef}) describes the effects appearing in the growth
direction at the expense of a simplified picture of the lateral
motion.

\begin{figure}[t]
{\epsfxsize=55mm
                \epsfbox[60 293 455 657]{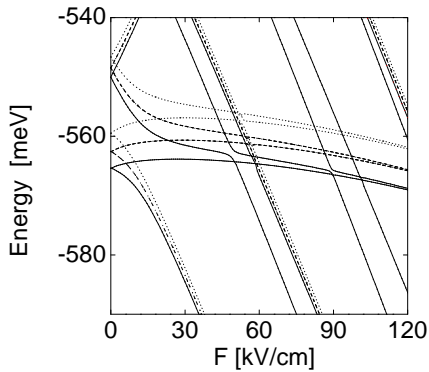}
}  \caption{The exact results solid lines and the results with the
frozen degree of freedom for equal hole and electron confinement
energies (dashed lines) and with equal hole and electron
confinement lengths (dotted line) for the parameters of Fig. 4
(b).  \label{comp}}
\end{figure}

Fig. 4 shows the comparison of the exact results (solid lines)
obtained with the separated center-of-mass and approximate results
calculated for frozen lateral degrees of freedom (dashed lines)
for identical quantum dots separated by a barrier of thickness
$b=4$ nm, as considered above in Fig. 4(b). The approximate method
reproduces the correct qualitative shape of the energy lines. Also
the recombination probability dependence on the electric field
does not significantly differ. However, the approximation of the
frozen lateral state eliminates the lateral excitations. The
avoided crossings of the bright energy levels with the dark energy
levels with lateral excitations are therefore overlooked in the
present approximation [cf. avoided crossing at $F=50$ kV/cm
missing for lines marked with dashed lines]. The accuracy of the
approximate method is better for dark states with separated charge
carriers than for the bright energy levels for which the electrons
and hole wave function overlap. \cite{1d} The discussed
approximation can be applied to evaluate the qualitative
dependence of the bright energy levels on the external field when
lateral excitations are absent. In the following section we will
use this approach to study the effect of the external field on the
negatively charged trion in coupled dots.

The dotted lines in Fig. 7 show the results of
frozen-degree-of-freedom calculations performed for the electron
confinement unchanged but weakened hole confinement for which the
lateral confinement radii of the electron and the hole are equal.
For weakened hole confinement the electron-hole interaction energy
is smaller, which leads to a blue-shift of the energy levels for
$F=0$ with respect to the equal confinement energies case (dashed
lines in Fig. 7). The interaction energy of the dissociated
electron-hole pair is less strongly affected by the change of the
hole localization strengths. Fig. 7 shows that the electric-field
dependence on the electric field is essentially not altered by the
strength of the hole localization, which justifies {\it a
posteriori} the assumption of the adopted center-of-mass
separability.

\section{stark effect for negative trion}

\begin{figure}[t]
\centerline{\hbox{
    \epsfxsize=55mm
                \epsfbox[41 152 452 500]
                {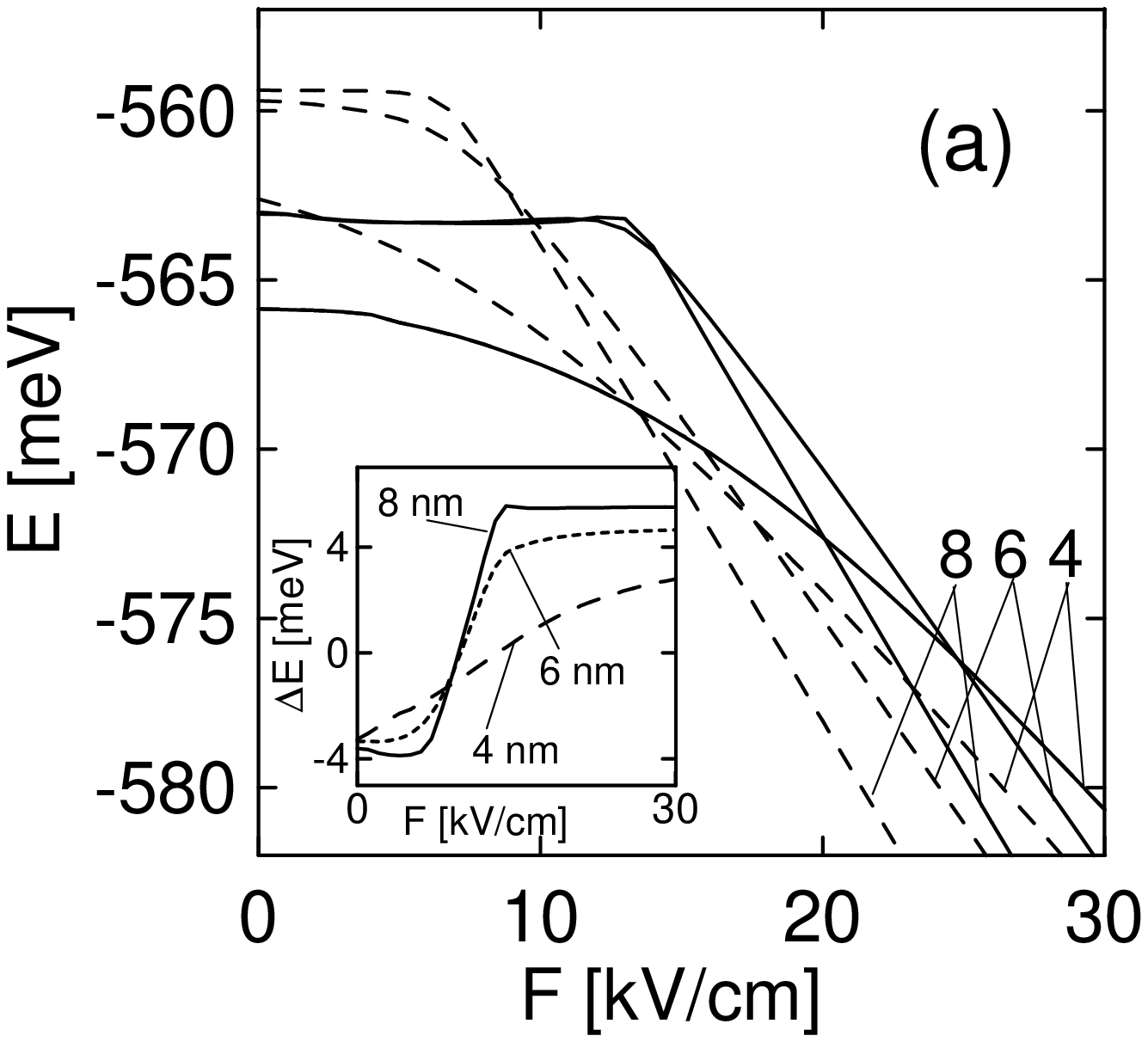}}}\vspace{0.2cm}
                \centerline{\hbox{
    {\epsfxsize=64mm\epsfbox[2 152 472 500]{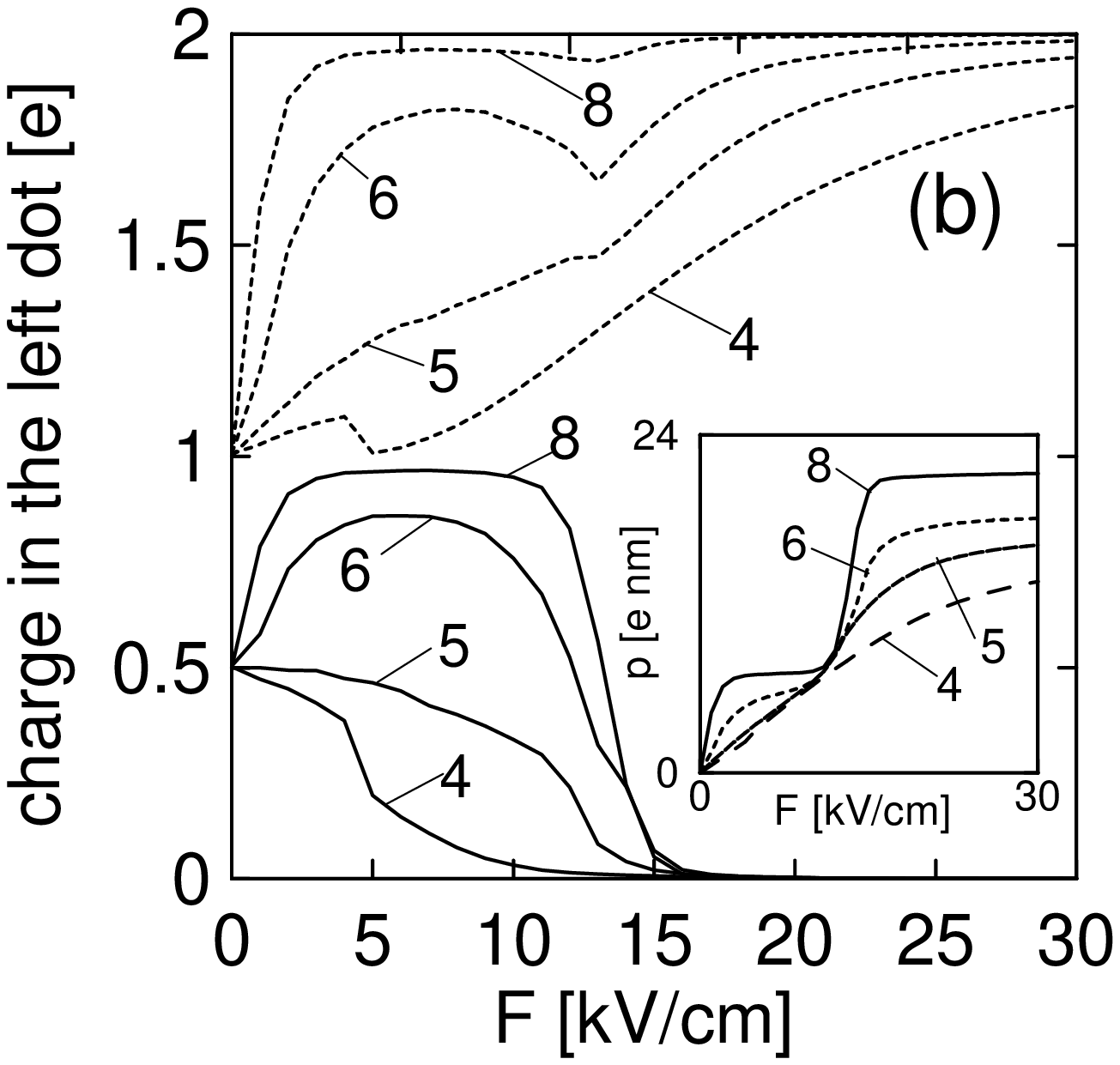}}}}
\caption{(a) Difference of the ground-state trion energy and the
electron ground state (trion recombination energy with respect to
GaAs energy gap - solid lines) and the exciton ground-state energy
(exciton recombination energy - dashed lines). The curves are
labelled by the barrier thickness $b$ in nanometers. Inset shows
the difference of the exciton and trion energy lines. (b) Electron
(dotted lines) and hole (solid lines) charge on the left side of
the origin as function of the electric field. Inset shows the
dipole moment.\label{trion1} }
\end{figure}

We consider the effect of the electric field on the ground state
of a negatively charged trion in which the electron subsystem is
in the singlet state. The approximation of the frozen lateral wave
functions will be used with the quasi one-dimensional\cite{1d}
electron-electron ($V_\mathrm{eff}^{ee}$) and electron-hole
interaction potential [Eq. (\ref{pot})]. Electron-electron
$V_\mathrm{eff}^{ee}$ potential is obtained for $m_h$ replaced by
$m_e$ in formula (\ref{pot}). The Hamiltonian for the trion reads
\begin{eqnarray} &H_{\mathrm{eff}}^{X-}=&
-\frac{\hbar^2}{2m_e}(\frac{\partial^2}{\partial z_{e1}^2}
+\frac{\partial^2}{\partial z_{e2}^2})
-\frac{\hbar^2}{2m_h}\frac{\partial^2}{\partial
z_h^2}+V_e(z_{e1})\nonumber \\ & &+V_e(z_{e2})+V_h(z_h)
-V_{\mathrm{eff}}(|z_{e1}-z_h|)\nonumber \\ &
&-V_{\mathrm{eff}}(|z_{e2}-z_h|)\nonumber
+V_{\mathrm{eff}}^{ee}(|z_{e2}-z_{e1}|) \\ & &
+e\Phi(z_{e1})+e\Phi(z_{e2})-e\Phi(z_h)+3\hbar\omega, \label{heft}
\end{eqnarray}
where $z_{e1}$ and $z_{e2}$ are coordinates of the first and
second electron, respectively.

Fig. \ref{trion1}(a) shows the difference of the trion ground
state energy and the ground state energy of a single electron as
function of the electric field for different values of the barrier
thickness and identical pair of quantum dots of width 6 nm. The
energy difference presented in Fig. \ref{trion1}(a) can be
identified\cite{ss} with the energy of the photon released when
the hole recombines with one of the electrons (calculated with
respect to the GaAs energy gap similarly as for the exciton). For
comparison the exciton ground state energy calculated with the
same approximation of the frozen lateral states is also shown by
the dashed lines. In the absence of the electric field the
recombination line of the negative trion has a lower energy than
the exciton recombination energy [cf. the inset to Fig.
\ref{trion1}(a)]. We found that the red-shift of the trion line is
smaller for smaller barrier thickness. This behavior as obtained
by neglecting the lateral correlations is in perfect qualitative
agreement with extensive variational calculations accounting for
both vertical and lateral correlations in a nearly exact
way.\cite{ss} Inset to Fig. \ref{trion1}(a) shows that for high
$F$ the energy difference of the trion and exiton energy lines is
an increasing function of $b$. This is due to the fact that the
interaction energy between the electrons confined in the same dot
is larger than the Coulomb interaction between the hole and
electron separated by the barrier.

For larger barrier thickness [cf. plots for $b=6$ and 8 nm in Fig.
8(a)] the recombination line of the trion is independent of the
electric field for $F$ lower than about 13 kV/cm. The flat part of
the plots corresponds to both the electrons and the hole staying
in the same quantum dot (as discussed above for exciton). We can
see that the ground-state of the trion is more resistant to the
dissociation by the electric field than the exciton ground state.
The exciton energy decreases faster than the trion recombination
line, which results in the reversal of the order of the lines at
$F=14$ kV/cm for $b=4$ nm and $F=10$ kV/cm for $b=6$ and 8 nm. For
large values of $F$ for which the hole and the electron charges in
both the exciton and the trion ground-states are completely
separated, the trion and exciton energy lines for each $b$ run
parallel to each other.

\begin{figure}[t]
\centerline{\hbox{
    \epsfxsize=40mm
                \epsfbox[210 480 575 1000] {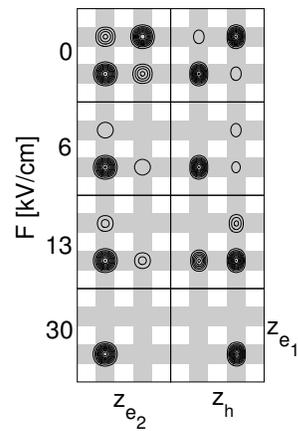}}
    }
\caption{Probability density integrated over the vertical
coordinate of one of the electrons (right panel) and the vertical
coordinate of the hole (left panel) for $b=6$ nm and different
values of the electric field $F$. The shaded areas show the
positions of the quantum dots. \label{trion2} }
\end{figure}

To explain the large stability of the trion ground state in the
symmetric coupled dots against dissociation by the electric field
we plotted in Fig. \ref{trion1}(b) the hole and the electron
charge accumulated in the left dot as a function of the electric
field for different barrier thicknesses. For large $b$ the
distribution of the electron and the hole charges between the dots
before the dissociation of the trion is qualitatively different
than in the exciton case (cf. Fig. 3). For $F=0$ the hole
(electron) charge in the right dot is 0.5 (1) due to the symmetry
of the system. For large barrier thickness ($b=8$ nm) the
electrons become localized in the left dot already under the
influence of a weak electric field. The hole initially follows the
electrons into the left dot (cf. the local maximum of the solid
line for $b=8$ nm). We remind the reader that for the exciton an
opposite behaviour was observed (cf. Fig. \ref{ce}): the electron
initially followed the hole for weak electric fields. The trion
becomes dissociated around 13 kV/cm, when the field moves the hole
from the left to the right dot. The reaction of the carriers on
the electric field is the most complex for $b=6$ nm [cf. Fig.
8(b)]. We have illustrated this in Fig. \ref{trion2} by additional
plots of the probability densities integrated over the vertical
coordinate of one the three particles. For zero electric field
there is a non-zero probability of finding the electrons in
different dots (cf. the left plot for $F=0$ in Fig. \ref{trion2}),
and the probability of finding an electron in a different quantum
dot than the hole (cf. the right plot for $F=0$ in Fig.
\ref{trion2}) is much smaller. For $b>8$ nm all the three
particles are found in the same dot. The leakage of particles to
the other dot seen in Fig. \ref{trion2} is a result of the
electron tunnel coupling which is already nonzero for $b=6$ nm. In
contrast to the case of $b=8$ nm, for $b=6$ nm a part of the
electron charge stays in the right dot when the field is switched
on (cf. the left plot for $F=6$ kV/cm in Fig. \ref{trion2}). When
the hole is transferred to the right dot (cf. the plots for $F=13$
kV/cm), part of the electron charge follows it, which results in a
local minimum of the electron charge accumulated in the left dot
for $F$ around 13 kV/cm [cf. Fig. \ref{trion1}(b) for $b=6$ nm].
For larger $F$ the particles become separated. For stronger tunnel
coupling between the dots, i.e. for $b=5$ and $4$ nm the hole
charge accumulated in the left dot depends on the external field
monotonically [cf. Fig. \ref{trion1}(b)], and a part of the
electron charge attempts to follow the hole when it leaves the
left dot. Therefore, for small $b$ the mechanism of the trion
resistance to dissociation becomes similar to the one observed for
the exciton (cf. Fig. \ref{ce}). The present results show that for
symmetric quantum dots the trion becomes dissociated into a pair
of electrons in one dot and the hole in the other without the
intermediate step consisting of an exciton confined in the right
dot and an electron in the left dot. This mechanism is more
clearly pronounced for larger $b$. The Coulomb interaction of the
electrons with the hole stabilizing the complex against
field-induced dissociation without the intermediate step is two
times larger than for exciton.

Note that for the trion [cf. Fig. \ref{trion1} (b)] the barrier
thickness has the opposite effect on the sensitivity of the
electrons and the hole to the electric-field induced localization.
For smaller $b$ the electric field is less effective in localizing
the electrons in the left dot but more effective in localizing the
hole in the right dot. The effect for the electrons is obviously
due to the strong electron tunnel coupling. For smaller $b$ a
smaller $F$ localizes the hole in the right dot because the energy
of its interaction with electrons changes less drastically after
the dissociation and the hole tunnel coupling is negligible.

The inset to Fig. \ref{trion1}(b) shows the electric dipole moment
for the trion as a function of the electric field. For high $F$,
when the particles are separated into different dots the dipole
moment takes the value $3e(w+b)/2$.\cite{przypis2} Note, that for
a thick barrier the dipole moment develops a plateau for the
region of fields in which the hole accompanies the electrons to
the right dot. For a thick barrier the recombination energy [cf.
inset of Fig. \ref{trion1}(a)] starts to change only when the
second plateau of the dipole moment is reached.

\begin{figure}[t]
\centerline{\hbox{
    \epsfxsize=50mm
                \epsfbox[5 152 436 500] {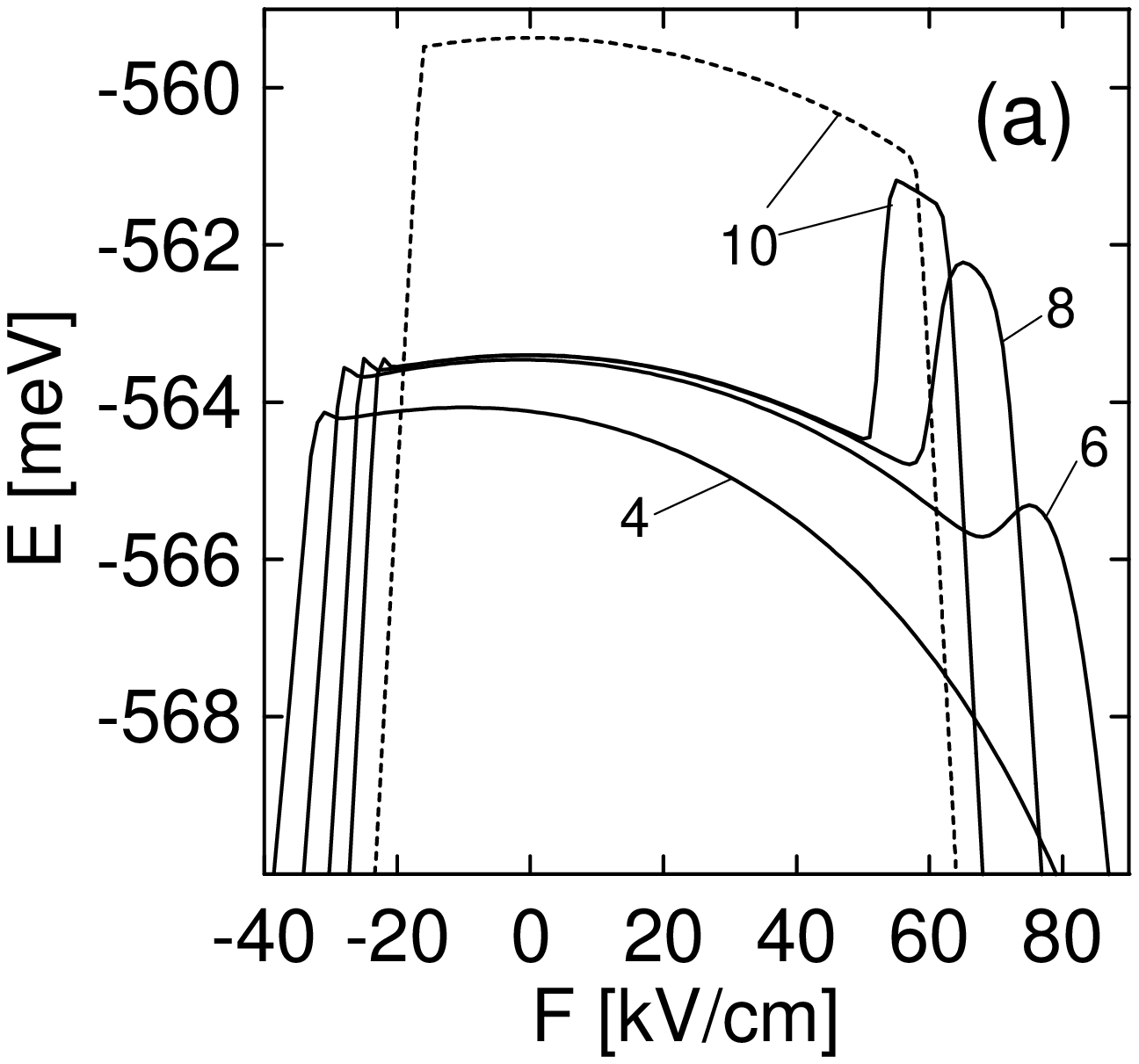}}
    {\epsfxsize=50mm\epsfbox[70 152 502 500]{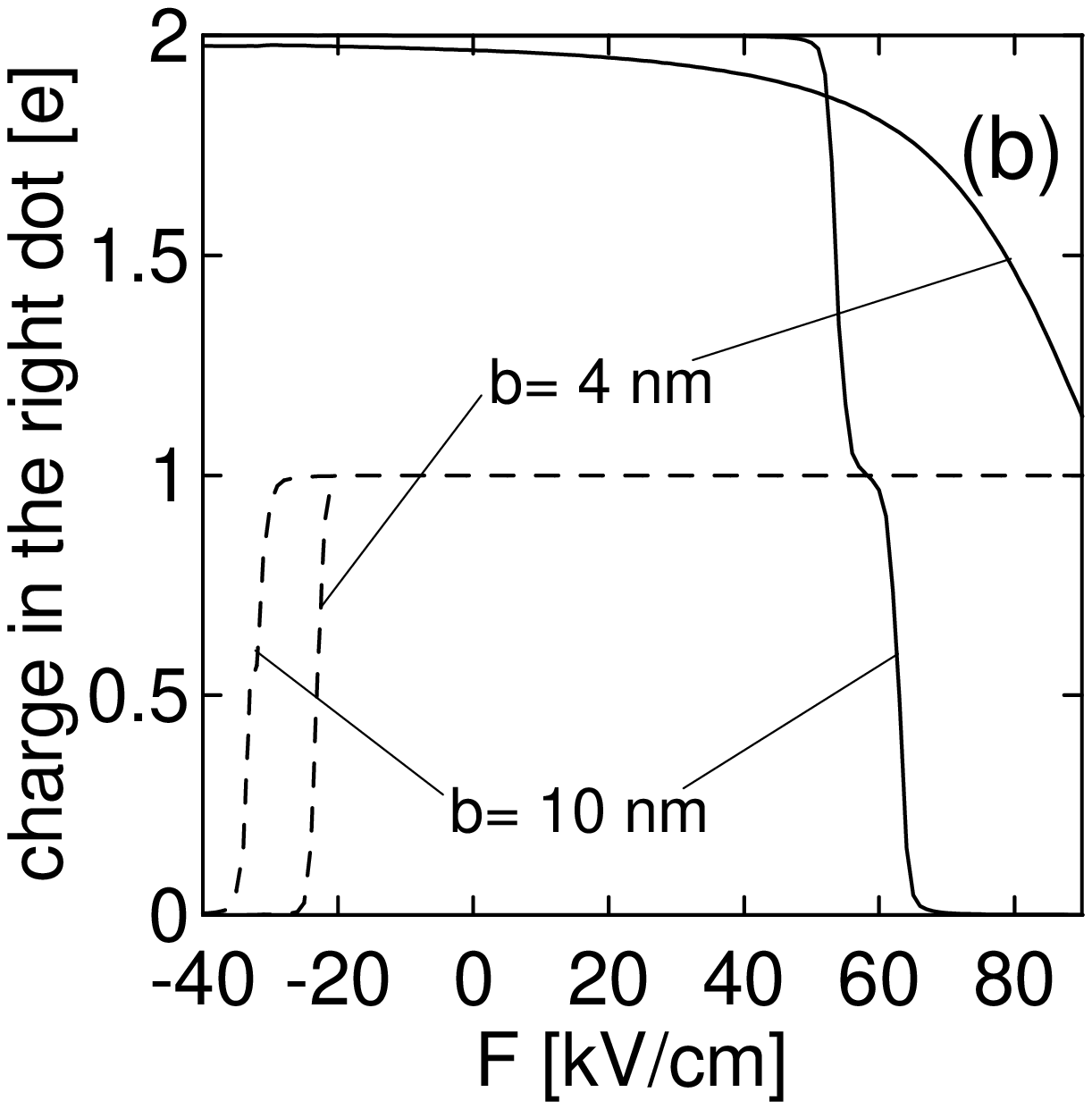}}
    }
\caption{(a) The trion recombination energies as functions of the
electric field (solid lines) for a pair of coupled quantum dots.
The curves are labelled by the barrier thickness in nanometers.
Right dot has a width of 6 nm and the width of the left dot is 4
nm. Dotted line shows the exciton recombination energy for $b=10$
nm. (b) The electron (solid lines) and the hole (dashed lines)
charge accumulated in the right dot for $b=4$ and 10 nm.
 \label{trionx} }
\end{figure}

We found a qualitatively different dissociation mechanism of the
trion in an asymmetric system of coupled dots. Suppose that the
right quantum dot has a thickness of 6 nm (as anywhere else in the
present paper) and that the left dot has a thickness of only 4 nm.
Fig. \ref{trionx}(a) shows the trion recombination energies for
different barrier thicknesses. The charge accumulated in the right
(wider) dot is plotted in Fig. \ref{trionx}(b). For $F=0$ the
three carriers stay in the right dot. For strongly coupled dots
($b=4$ nm) the electrons resist strongly to being removed to the
thinnest dot. For $F=90$ kV/cm less than one elementary charge is
localized in the left dot. On the other hand the negative electric
field removes abruptly the hole to the thinnest dot at $F<-25$
kV/cm. For thicker interdot barrier the trion recombination energy
develops a local maximum for positive electric fields [see the
plots for $b=6$, 8 and 10 nm in Fig. \ref{trionx}(a)]. Let us
analyze the origin of these maxima for the case of $b=10$ nm. For
positive electric field up to 50 kV/cm both the electrons are
confined in the right dot [cf. Fig. 10(b)]. Then between $F=50$
kV/cm and $F=55$ kV/cm one of the electrons is transferred to the
right dot. In this electric field range the trion is dissociated
into an exciton confined in the right dot and a spectator electron
in the left quantum dot. The final state after the trion
recombination, i.e. the ground state of a single electron, is
localized in the left quantum dot for $F>50$ kV/cm, i.e. for the
same value of the electric field which induces the transition of
the first electron from the trion state to the left quantum
dot.\cite{przypis} After the trion dissociation the recombination
energy almost reaches the recombination energy of the exciton [cf.
dotted line in Fig. \ref{trionx} (a)]. The slight redshift of the
dissociated trion line with respect to the exciton in this
electric field range is due to the Coulomb perturbation of the
exciton remaining in the right quantum dot by the spectator
electron in the left quantum dot. The second electron is removed
from the right dot between 60 and 65 kV/cm.

Note, that the observed mechanism of dissociation of the trion
into an exciton and an electron does not occur in the system of
symmetric quantum dots (cf. Fig. \ref{trion1}). For asymmetric
quantum dots the stronger confinement energy in the thinner of the
dots prevents the second electron from entering it simultaneously
with the first one. In the asymmetric system the exciton becomes
dissociated into an electron and a hole for {\it larger} electric
fields than the one inducing dissociation of the trion into an
exciton and a free electron [cf. Fig. \ref{trionx}(b)]. On the
other hand the exciton created in the right quantum dot after the
trion dissociation is more resistant to the electric field induced
dissociation than the exciton. The electron remaining in the right
dot is less willing to pass to the left quantum dot if it is
already occupied by an electron.

The recombination energy lines of the trion in the asymmetric
system of coupled dots present a positive second derivative with
respect to the electric field for a certain range of $F$. Namely,
for $b=6$ nm the second derivative is positive for the electric
field range $F\in(61,71)$, $F\in(57,60)$ and $F\in(50,52)$ kV/cm
for $b=6$, 8 and 10 nm, respectively.

For symmetric dots the mechanism behind the trion dissociation
into an electron pair confined in one dot and the hole in the
other, without an intermediate step consisting of an exciton in
one dot, and the electron occupying the other quantum dot, is
easily explained when considering large $b$ using a simple
reasoning in which the tunnel effect and the interdot Coulomb
interactions are neglected. In this model the dependence of the
energy of the trion on the external field can be written as
$E_{X^-}=-2E_{eh}+E_{ee}-Fbe/2$, where $E_{eh}$ ($E_{ee}$) is the
absolute value of the electron-hole interaction for the particles
localized in the same dot. The trion is localized in the dot in
which the electron localization is favored by the field. The
lowest energy level corresponding to the exciton confined in one
dot and the electron in the other is $E_X=-E_{eh}-Fbe/2$, and the
energy level corresponding to a completely dissociated system is
$E_d=E_{ee}-3Fbe/2$. For $F=0$ the trion is bound for
$E_{eh}<E_{ee}$ and the ground state energy equals either
$E_{X^-}$ or $E_d$. The energy splitting of $E_{X^-}$ and $E_X$ is
not affected by the field which explains the absence of an exciton
as an intermediate step of trion ionization. A similar simple
reasoning can be used for coupled asymmetric dots where the
intermediate step of trion dissociation is now found.

\section{Discussion}

As mentioned in the introduction previous ground-state
calculations\cite{Leburton,Karen} for the Stark effect in
vertically coupled dots detected a deviation of the energy
dependence on the electric field from the expected quadratic form
obtained within the second order perturbation theory.\cite{TCK}
The inset to Fig. 1(a) shows that for identical quantum dots this
deviation, i.e., a cusp of the recombination energy in function of
the electric field, is due to a narrow avoided crossing of two
lowest energy levels. In the absence of the electric field these
two energy levels are nearly degenerate. This near degeneracy
results from the smallness of the hole tunnel coupling between the
dots. For the case presented in Fig. 1(a) these two energy levels
correspond to opposite parity of the hole [cf. Fig. 2(a) for
$F=0$]. The electric field easily mixes the two energy levels
localizing the hole in the right dot (in the ground state) and in
the left quantum dot (in the first excited state). When the
confinement potential is asymmetric the discussed anticrossing of
the two lowest energy levels are replaced by a crossing (cf. Fig.
6). This is due to a nearly complete localization of the hole in
left or right quantum dot in the two states. The cusp of the
ground-state is produced by {\it two} energy levels crossing or
nearly crossing. It is therefore clear that second order
perturbation theory for a single {\it nondegenerate} energy level
given\cite{TCK} for a {\it single} quantum dot is not applicable
to the ground-state in coupled quantum dots. There is therefore no
reason for which the ground-state should follow the quadratic
formula and the deviation from parabolicity does not really
deserve to be called an anomaly.

In the present paper we have found another deviation from the
common quadratic Stark shift, also involving two energy levels.
This deviation appears for an intermediate barrier thickness and
is due to an avoided crossings of a bright energy level with both
carriers in the same dot and a dark energy with separated charge
carriers. This unusual Stark effect, shown in Fig. 2(c) for a
symmetric dot, should be visible in low-excitation PL
spectroscopy.\cite{Bayer} The observation of the excited exciton
states should be facilitated by a relatively weak tunnel coupling
between the quantum dots. In the corresponding PL spectrum, one of
the lines should be independent of the electric field in both
energy and intensity. The additional structure below and above the
constant-energy line should be observed in the form of an
anticrossing. The intensity of the constant-energy line should be
reduced in the region, in which the anticrossing appears.

Real InAs/GaAs quantum dots exhibit a strain-induced intrinsic
dipole moment at $F=0$.\cite{Fry} The intrinsic dipole moment has
been neglected in the present calculations. However, the unusual
Stark shift for the coupled dots is predicted for quite small
electric fields (lower than 15 kV/cm), for which the effect of the
intrinsic dipole moment is negligible. For comparison in the
experiment the intrinsic dipole moment leads to a shift of the
transition energy by about 5 meV for $F=100$ kV/cm.\cite{Fry}
Therefore, the intrinsic dipole moment should not modify the
qualitative features of the effect predicted in the present paper.
The second order effect of the polarizability related to the
electric-field induced deformation of the electron and hole wave
function for the discussed low electric field range should be even
smaller. Similar mechanism of the exciton dissociation via an
avoided crossing has been found for asymmetric dots [cf. Figs.
5(a) and (b)]. The difference between the ideally symmetric system
and the more realistic asymmetric one is that the bright state
which does not participate in the avoided crossing is shifted to a
different energy, lower or higher depending on the direction of
the electric field. The mechanism of the exciton dissociation via
an avoided crossing of a dark and a bright energy level described
here has been recently confirmed experimentally\cite{nexp} after
the present paper has been submitted.

Second-order perturbation theory for a single nondegenerate energy
level\cite{TCK} predicts a nonpositive curvature of the energy
level as a function of the electric field. Although the curvature
is indeed nonpositive in the ground-state, a positive curvature is
obtained for the excited bright energy levels in the presence of
the avoided crossings with lower energy levels [see Figs. 4(a-c),
9 and 11]. A non-degenerate perturbation theory for a single
level\cite{TCK} obviously does not apply for the the energy level
interaction.

In view of the present results the pronounced drop of the
recombination energy for a bias voltage for which an electron is
trapped in the quantum dot\cite{Ruth} closer to the electron
reservoir can be understood provided that the recombination signal
in the observed range of wavelengths comes from this dot.
Otherwise, the charge of the electron trapped in the dot closer to
the reservoir would have a negligible influence on the energy of
exciton recombination in the other dot separated by a barrier of
12 nm [cf. the small energy spacing between the exciton
recombination lines with and without a spectator electron in the
other dot for $b=10$ nm in Fig. 10(a)]. The drop would result from
the electrostatics of the negative trion in which the energy of
the electron-hole attraction is larger than the electron-electron
repulsion due to a difference of the strength of lateral
localization of the carriers (see the discussion given in Ref.
[\cite{ss}]). The observed growth of the recombination energy for
the smaller absolute value of the bias voltage could be related to
a passage of one of the electrons to the upper dot. The presented
calculations for the trion were limited to the ground-state.
However, the PL line observed in the experiment which we here
attribute to the trion recombination in the {\it lower} of the
dots does not correspond to the ground state since the quantum dot
in the upper layer are {\it larger}. Therefore, in the experiment
the dissociation of the trion localized in the lower dot could be
associated with an avoided crossings with lower energy states,
which as obtained for the exciton, can produce a positive
curvature of the recombination line over a wide range of electric
field values.

\section{Summary and Conclusions}

We have studied the exciton and negative trion states in a simple
but exactly solvable model of vertically coupled quantum dots
allowing for a description of the effects related to the
modification of the electron-hole interaction by an electric field
applied in the growth direction. The effect of the tunnel coupling
between the dots and the confinement potential asymmetry was
considered. The mechanism of the electric-field induced exciton
and trion dissociation was described.

We have shown that the previously\cite{Leburton,Karen} found
deviations from the quadratic Stark effect are due to energy
levels crossings (or very narrow avoided crossings). For weaker
tunnel coupling we have found another non-quadratic feature due to
an avoided crossing of bright and dark energy levels. This feature
appears also in the presence of the asymmetry of the coupled dots
and is due to the Coulomb interaction. Positive curvature of the
bright excited exciton energy levels is obtained in the range of
electric fields corresponding to avoided crossings with lower
levels.

Although in the presence of asymmetry of the coupled dots the
trion is dissociated into an exciton and an electron by the
electric field, for symmetric dots the dissociation mechanism is
different, i.e. the trion is directly separated into an electron
pair in one dot and the hole in the other. The trion is  more
stable against this mechanism of dissociation than the exciton.
The process of trion dissociation into an exciton and a free
electron that we obtain for the case of asymmetric coupling leads
to a positive curvature of the PL line as a function of the
electric field which has never been observed for the exciton
ground state.

{\bf Acknowledgments} This paper was supported by the Polish
Ministry of Scientific Research and Information Technology in the
framework of the solicited grant PBZ-MIN-008/P03/2003, the Flemish
Science Foundation (FWO-Vl), the Belgian Science Policy, the
University of Antwerpen (VIS and GOA) and the European Commission
GROWTH programme NANOMAT project, Contract No. G5RD-CT-2001-00545,
and the EU-NoE SANDiE. T.C. is partly supported by the Marie Curie
Training Site Programme of the European Union. B.S. is supported
by the Foundation for Polish Science (FNP) and by the EC Marie
Curie IEF project MEIF-CT-2004-500157. We are thankful for Dr.
Ruth Oulton for helpful discussions and for making the
experimental data accessible prior to publication.
\newline


\begin{thebibliography}{00}
\bibitem{przypis2} For the trion, unlike for the exciton, the value of the dipole moment
depends on the choice of the origin. The values displayed in the
inset of Fig. 8 (b) are calculated for the origin set in the
middle of the barrier.
\bibitem{przypis}  If the ground state of a single electron was localized in the left dot
for higher value of the electric field, the recombination energy,
calculated as the difference of the ground state of the trion and
electron, would have a doubtful physical interpretation since it
would additionally assume that the recombination of the
electron-hole pair is accompanied by the passage of the electron
from the right to the left dot.
\end{thebibliography}
\end{document}